\documentclass[fleqn,usenatbib]{mnras}

\usepackage{newtxtext,newtxmath}
\usepackage{amsmath}
\usepackage[T1]{fontenc}
\usepackage{ae,aecompl}

\usepackage{appendix}
\usepackage{graphicx}% Include figure files
\usepackage{dcolumn}% Align table columns on decimal point
\usepackage{bm}% bold math
\usepackage{aas_macros}
\usepackage{siunitx}	% proper SI notation
\usepackage{cleveref}	% adds cref command
\usepackage{enumitem}
\usepackage{csquotes}
\usepackage[all]{nowidow}		% tries to avoid widows
\DeclareSIUnit \c 			{\ensuremath{\mathit{c}}}
\DeclareSIUnit \parsec 		{pc}
\DeclareSIUnit \eV 			{eV}
\DeclareSIUnit \eVcc		{eV\per\c\squared}
\DeclareSIUnit \Msun 		{M_\odot}

\newcommand\ddfrac[2]{\frac{\displaystyle #1}{\displaystyle #2}}

\usepackage{todonotes}

\title[On the formation and stability of fermionic dark matter halos in a cosmological framework]{On the formation and stability of fermionic dark matter halos in a cosmological framework}

\author[C. R. Argüelles et. al.]{
Carlos R. Argüelles,$^{1}$\thanks{E-mail: carguelles@fcaglp.unlp.edu.ar}
Manuel I. Díaz,$^{2,3}$\thanks{E-mail: manuel.diaz@phys.ens.fr}
Andreas Krut,\thanks{E-mail: andreas.krut@gmx.de}
and Rafael Yunis$^{4,5}$\thanks{E-mail: yunis121@gmail.com}
\\
$^{1}$Facultad de Ciencias Astron\'omicas y Geof\'isicas, Universidad Nacional de La Plata, Paseo del Bosque, B1900FWA La Plata, Argentina\\
$^{2}$Laboratoire de Physique, \'Ecole Normale Sup\'erieure, CNRS, Universit\'e PSL, Sorbonne Universit\'e, Universit\'e de Paris, 75005 Paris, France\\
$^{3}$Departamento de Física, Facultad de Ciencias Exactas y Naturales, Universidad de Buenos Aires, Pabellón I, Ciudad Universitaria, 1428 Buenos Aires, Argentina. \\
$^{4}$ ICRANet, Piazza della Repubblica 10, I--65122 Pescara, Italy.\\
$^{5}$ Physics Department, La Sapienza University of Rome, P.le Aldo Moro 5, I--00185 Rome, Italy.
}

\date{Accepted XXX. Received YYY; in original form ZZZ}

\pubyear{2020}

\begin{document}
\label{firstpage}
\pagerange{\pageref{firstpage}--\pageref{lastpage}}
\maketitle

\begin{abstract}
The formation and stability of collisionless self-gravitating systems are long-standing problems, which date back to the work of D. Lynden-Bell on violent relaxation and extends to the issue of virialization of dark matter (DM) halos. An important prediction of such a relaxation process is that spherical equilibrium states can be described by a Fermi-Dirac phase-space distribution, when the extremization of a coarse-grained entropy is reached. In the case of DM fermions, the most general solution develops a degenerate compact core surrounded by a diluted halo. As shown recently, the latter is able to explain the galaxy rotation curves while the DM core can mimic the central black hole. A yet open problem is whether this kind of astrophysical \textit{core-halo} configurations can form at all, and if they remain stable within cosmological timescales. We assess these issues by performing a thermodynamic stability analysis in the microcanonical ensemble for solutions with given particle number at halo virialization in a cosmological framework. For the first time we demonstrate that the above \textit{core-halo} DM profiles are stable (i.e. maxima of entropy) and extremely long lived. We find the existence of a critical point at the onset of instability of the \textit{core-halo} solutions, where the fermion-core collapses towards a supermassive black hole. For particle masses in the \si{\kilo\eV} range, the core-collapse can only occur for  $M_{\rm vir} \gtrsim \SI{E9}{\Msun}$ starting at $z_{\rm vir}\approx 10$ in the given cosmological framework. Our results prove that DM halos with a \textit{core-halo} morphology are a very plausible outcome within nonlinear stages of structure formation.

% The latter is able to explain the galaxy rotation curves while the DM core can mimic the central black hole
%  and due to the Pauli exclusion principle
% able to explain the galaxy rotation curves
% --- composed either by classical (e.g. stars) or quantum (e.g fermions) particles ---
%The full methodology here adopted is: i) we calculate the linear power spectrum for the fermionic candidates; ii) we use the corresponding Press-Schechter formalism to obtain the virial halo mass, $M_{\rm vir}$, with associated redshift $z_{\rm vir}$; iii) by assuming a maximum entropy production principle for halo formation,  we obtain the full family (from King-like to  \textit{core-halo}-like) of such fermionic DM profiles in agreement with above virial constraints; iv) we calculate the stability and typical life-time of such equilibrium states within a thermodynamic approach.
%We assess these issues by performing a thermodynamic stability analysis in the microcanonical ensemble for solutions with given particle number $N$ at the moment of halo-virialization in cosmology, and further calculate the life-time of the equilibrium states
% Importantly, this statement is only true for tidally truncated systems but does not hold for (artificially) box-confined ones.
%  Therefore they can be reachable in Nature
%  When the total particle number is above the Oppenheimer-Volkoff value ($N>N_{\rm OV}$), there exist both stable and unstable configuration-branches, being the central core in the latter branch able to collapse towards a supermassive black hole.
\end{abstract}
%  when $m \sim 10\,\si{\kilo\eV}$


%%%%%%%%%%%%%%%%%%%%%%%%%%%%%%%%%%%%%%%%%%%%%%%%%%

%%%%%%%%%%%%%%%%% BODY OF PAPER %%%%%%%%%%%%%%%%%%

\section{Introduction}
\label{sec:introduction}

The thermodynamics of self-gravitating systems is a vast subject of research with deep implications in astrophysics. It includes the long standing problems of relaxation in collisionless/collisional stellar systems, to the issue of virialization of DM halos and structure formation. Depending on the initial conditions of the system at virialization (such as total mass, size, degree of concentration, etc.) they may become thermodynamically unstable, experience a gravothermal catastrophe, suffer phase-transitions or even undergo core-collapse towards a massive black hole (BH).

The first dynamical and thermodynamical studies were developed for classical point masses within Newtonian gravity, and applied to the case of stars in globular clusters or galaxies in \citep{1962spss.book.....A,1963MNRAS.125..127M,1966AJ.....71...64K,1967MNRAS.136..101L,1968MNRAS.138..495L,1980MNRAS.190..497K,1990PhR...188..285P,Katz2003,2006IJMPB..20.3113C}.\footnote{
	In late '70s a more rigorous approach was developed \citep{katz1978number} to analyze the thermodynamic stability from statistical mechanics, and applied here for the case of fermions.
} Contemporaneously to such earliest works, the attention was directed to self-gravitating quantum particles in \citet{1969PhRv..187.1767R}, mainly focusing in the mathematical aspects of the solutions at hydrostatic-equilibrium. The role of the quantum pressure at the center of the configurations was there evidenced within a fully general relativistic (GR) treatment, distinguishing between the bosonic and fermionic nature of the particles. In the case of fermions, such self-gravitating models (either at zero or at the more realistic finite temperature) were further investigated by different authors, with applications to the morphology of DM halos at equilibrium, and/or to the mass of the DM particles \citep{1983PhLB..122..221B,1983A&A...119...35R,1990A&A...235....1G,1992A&A...258..223I,2002PrPNP..48..291B,2003LNP...616...24B,2014JKPS...65..801A,2014MNRAS.442.2717D,2015MNRAS.451..622R,2015PhRvD..91f3531C,2015PhRvD..92l3527C,2016PhRvD..94l3004G,2017MNRAS.467.1515R,2018PDU....21...82A,2019PDU....24..278A,2019IJMPD..2843003A,2020A&A...641A..34B}, though lacking a thermodynamic-stability analysis (with the exception of \citealp{2015PhRvD..91f3531C,2015PhRvD..92l3527C}).

In the case of galactic structures, a central question in the field is precisely how a self-gravitating system of collisionless particles (either stars or DM particles) reaches the steady state we observe. The complex evolution of the coarse-grained phase-space distribution of such a system is driven by the mean-field Vlasov-Poisson (VP) equation. It involves rapid fluctuations in the overall time-varying gravitational field, able to redistribute the energy between the particles in few dynamical times, even faster than collisional relaxation-time scales \citep{2008gady.book.....B}. Such a process is known as violent relaxation, and is the main mechanism able to lead the averaged phase-space distribution function (DF) towards a quasi stationary state (qSS).

A typical stationary coarse grained DF which is a possible solution of the mean-field VP equation was originally predicted to be of a Fermi-Dirac type in the seminal work of D. Lynden-Bell in \cite{1967MNRAS.136..101L}, from a statistical description. Such an approach for violent relaxation was developed for distinguishable classical particles, and three decades later extended to indistinguishable particles such as massive neutrinos in \citet{1996ApJ...466L...1K}. Those results were later formalized and extended out of equilibrium (e.g. allowing for escape of particle effects) within thermodynamical and kinetic theory approaches in \citet{1998MNRAS.300..981C} and \citet{2004PhyA..332...89C} respectively. Such a generalization leads to a tidally truncated Fermi-Dirac DF at equilibrium, naturally implying finite sized systems as the ones applied here to model fermionic halos.

However, as recognized by Lynden-Bell himself \citep{1967MNRAS.136..101L}, the violent relaxation mechanism is likely to be incomplete towards the outer and low density halo regions. This discrepancy was traditionally associated to the short dynamical time-scales involved, which together with the long range nature of the interactions, imply that the entire system has not enough time to explore the full phase-space to reach a most likely final state. Contrary to this expectation, in \citet{2008PhRvE..78b1130L} it was shown that when the initial phase-space distribution (prior to relaxation) satisfy the virial theorem, then the resulting qSS is close to ergodic and so the Lynden-Bell statistics works well to reproduce the numerical simulations.\footnote{
  An effective cut-off in radius is needed in order for this virial condition scenario to match the simulations \citet{2008PhRvE..78b1130L}.
} Though, if the virial condition is violated, parametric resonances arise and the mean-field gravitational potential oscillations lead to ergodicity breaking with partial mass evaporation towards the halo \citep{2008PhRvE..78b1130L}. This more general case implies a qSS with a \textit{dense core}--\textit{diluted halo} behavior, as arising within numerical simulations, which according to \citet{2014PhR...535....1L} it can not be well explained in the traditional Lynden-Bell theory. However, more general \textit{core-halo} distributions can arise as well within generalized Lynden-Bell statistics (as e.g. successfully applied here for fermionic halos with \cref{eq:FD-DF-cutoff}), which still remain to be contrasted against numerical simulations of violent relaxation.
% (see \citet{2014PhR...535....1L} for a more general discussion)

%ADD HERE A FULL PARAGRAPH ABOUT GENERAL qSS ARISING FROM GENERAL (NON-FULLY ERGODIC) LONG-RANGE INTERACTING SYSTEMS, AND MENTION UNDER WHICH CONDITIONS CAN LB THEORY BE APPLIED TO SELF-GRAVITATING SYSTEMS (citing Levin et al. works), though an yet open issue for DM halos made of self-gravitating fermions, which we are going to shed light here as well!!

On general grounds, as first recognized in the original work of D. Lynden-Bell \citep{1967MNRAS.136..101L} for classical particles, there is an effective \enquote{exclusion principle} acting in such relaxation processes due to the incompressibility of the VP equation in phase-space. That is, the coarse-grained DF cannot exceed a maximum initial value after some evolution. This may be considered as a classical counterpart of the Pauli exclusion principle for self-gravitating fermions, both leading to degeneracy effects in the matter distribution of the collisionless systems. This degeneracy, either of quantum or classical origin, when large enough implies a richer  \textit{dense~core}--\textit{diluted~halo} morphology for the qSS (regardless any possible incompleteness of violent relaxation which would only affect the outer halo). Explicit realizations of such a novel profile morphology was solved in \citet{1998MNRAS.296..569C} for stars, and \citet{1990A&A...235....1G,2002PrPNP..48..291B,2015MNRAS.451..622R,2015PhRvD..92l3527C,2018PDU....21...82A,2019PDU....24..278A,2020A&A...641A..34B} for fermions. Such distributions differentiate with respect to the traditional King profiles (typically applied to globular clusters), being the later based on a Boltzmannian DF inherent to collisional systems \citep{2008gady.book.....B}. 
% \footnote{
  % For the case of the Milky Way, in \citet{2020MNRAS.498.5574E} it was recently shown that the equilibrium assumption for the DM halo breaks above $\sim 30$ kpc, implying the DM halo is not relaxed only above that scale.
 %}

The manifestation of such a \textit{core-halo} profile in astrophysics is yet an open issue, though several key results have been reported from theoretical as well as from phenomenological fronts. They include (i) the avoidance of the traditional gravothermal catastrophe, thanks to the arising of the central degeneracy \citep{1998MNRAS.296..569C,2015PhRvD..92l3527C}, not present in Boltzmannian distributions; (ii) the possibility for the degenerate fermion-core to mimic the massive BHs at the center of galaxies, while the outer halo can explain the rotation curves \citep{2015MNRAS.451..622R,2018PDU....21...82A,2019PDU....24..278A}; (iii) the fact that astrophysical \textit{core-halo} DM profiles can be thermodynamically (and dynamically) stable, as well as long lived as proven here. These issues can be addressed through a thermodynamic stability analysis for self-gravitating systems with given particle number (see e.g. \citealp{2015PhRvD..92l3527C} for a recent study within Newtonian gravity). Indeed, it is the aim of this work to make, for the first time, such an analysis in full GR, and apply it to realistic DM halos of fermionic nature at virialization within a realistic cosmological framework.

On the realm of fermionic self-gravitating systems, the bulk of the works in the field (besides the ones done by P. H. Chavanis) were developed assuming hydrostatic-equilibrium, without analyzing the thermodynamic stability of the solutions. Interestingly, stationary phase-space solutions of the VP equation (of the form $f(\epsilon)$ with $f'(\epsilon)<0$ and $\epsilon$ the particles energy) such as (\ref{eq:FD-DF}) or (\ref{eq:FD-DF-cutoff}) implemented here, are always VP dynamically stable \citep{2008gady.book.....B,2011CMaPh.302..161L}. Though, this stability analysis does not explain how such a collisionless self-gravitating system reaches the required steady state, nor if they minimize the free energy (or maximize the coarse-grained entropy), or if they are just transitional (unstable/unreachable) states. 

Indeed, a proper answer to this problem requires the introduction of relaxation processes such as of violent-relaxation and Landau damping, where the qSS is reached upon a maximization (coarse-grained) entropy problem. Moreover, as it is explained in \citet{2015PhRvD..92l3527C}, the VP equation has an infinite number of conserved integrals (e.g. the so called Casimir integrals), while the maximization entropy problem holds only for given total mass and energy (two integrals). Therefore VP dynamical-stability \textit{does not} necessarily imply thermodynamical stability (see however \citealp{2019arXiv190810806C} and references therein for general conclusions in the relativistic case). Thus some VP dynamically-stable solutions in hydrostatic equilibrium are more likely to occur in Nature than others. 
% \footnote{
    %See appendix A in \citet{2015PhRvD..92l3527C}, for a more detailed discussion in the case of fermionic phase-space distributions including for particle evaporation effects.
%}

A first thermodynamic study of self-gravitating systems of elementary fermions in a full relativistic framework was given in \citet{1999EPJC...11..173B}, paying attention mainly to the occurrence of gravitational phase-transitions between gaseous and semi-degenerate (neutral) fermion stars. They worked in the canonical ensemble, and analyzed only (i) systems with total number of fermions below the Oppenheimer-Volkoff value (i.e. $N<N_{\rm OV}$) where no relativistic collapse to a BH is possible, and (ii) very low spherical-box sizes $R$ up to $R \sim 10^1 R_{\rm OV}$ not applicable to any astrophysical DM halo system. 

Only recently, a more extensive full relativistic, thermodynamic stability analysis for self-gravitating fermions was done in \citet{2019CQGra..36f5001R,2019arXiv190810303C,2020EPJB...93..208A}, complementing to the case $N>N_{\rm OV}$ (where core-collapse towards a BH may arise). In the latter, such an analysis was done either in the microcanonical or canonical ensembles, and enlarging to spherical-box sizes about an order of magnitude above the original work of \citet{1999EPJC...11..173B}, yet orders of magnitude below any realistic DM halo size. An analogous non-relativistic (e.g. Newtonian) thermodynamical study was done in \citet{2015PhRvD..92l3527C} within the more realistic fermionic King model. There, much larger bounded system sizes were reached, with potential applicability to realistic DM halos for particles masses $mc^2 \sim \SI{1}{\kilo\eV}$ (results which are here compared against our relativistic approach in \cref{sec:comparison}). Though, the works cited above were mainly focused in the mathematical aspects and characterization of the solutions in full dimensionless units, and only applied to astrophysical objects in a qualitative fashion.

It is the purpose of this work to make a detailed thermodynamic stability analysis of a self-gravitating system of fermions in GR, at the moment of DM halo formation in a realistic cosmological setup, and for particle masses of $mc^2 \sim \mathcal{O}(\SI{10}{\kilo\eV})$. The novelty of this work with respect to other similar analysis recently given in the literature \citep{2015PhRvD..92l3527C,2019CQGra..36f5001R,2020EPJB...93..208A}, relies in the fact it is the first time such a thermodynamical analysis of fermionic matter is developed in GR, for realistic DM halo configurations which are tidally-truncated (i.e. by including for escape of particle effects) in a Warm Dark Matter (WDM) cosmology. In particular, we emphasize on the possible different shapes of the DM profiles, and analyze their thermodynamical stability, from dwarf to larger halo sizes, making a direct link with the \textit{core-halo} Ruffini-Argüelles-Rueda (RAR) profiles presented in \citet{2015MNRAS.451..622R} and in \citet{2018PDU....21...82A,2019PDU....24..278A,2020A&A...641A..34B}.

In the following, we summarize the content of the present work. In \cref{sec:equilibrium-eqtns} we write the hydrostatic and thermodynamic equilibrium equations of a self-gravitating system of massive fermions at finite temperature in GR.
In \cref{sec:thermo-stability-analysis} we develop for the first time a thermodynamic stability analysis for systems with given particle number $N$ at halo virialization (with initial conditions consistent with a Press-Schechter analysis within a WDM cosmology as given in \cref{sec:appendix:halo-formation}), and for a particle mass of  $mc^2 \sim \mathcal{O}(\SI{10}{\kilo\eV})$. We first assume: a) spherical-box confined configurations with size $R$; and b) tidally truncated systems. We work in the microcanonical ensemble, which is the more appropriate for astrophysical applications as explained in \citet{2015PhRvD..92l3527C}. Details on the stability criterion used here, as well as the calculation of the life-time of metastable states are given in \cref{sec:appendix:stability_examples}. 

The main conclusion of \cref{sec:thermo-stability-analysis} can be summarized as follows: \textit{for spherical box-confined configurations, there are no thermodynamically stable core-halo solutions which are of astrophysical interest when applied to DM halos, while the opposite is true for tidally truncated systems}. This is the first time that a \textit{dense~core}--\textit{diluted~halo} solution, which is successfully applied to model realistic DM halos (as in \citealp{2018PDU....21...82A,2019PDU....24..278A,2020A&A...641A..34B}), is proven to be thermodynamically stable (i.e. entropy maxima) and extremely long-lived.

In \cref{sec:comparison} and \cref{sec:appendix:stability_examples} we compare our results with other similar works in the literature. In \cref{sec:TurningPoint} we explain the mathematical and physical characteristics of the \textit{turning-point} instability, while further show the existence of a gravitational core-collapse for solutions with $N > N_{\rm OV}$, and its implications for super massive BH formation in the high redshift Universe. In \cref{sec:conclusion} we draw our conclusions.

%In \cref{sec:appendix:stability_examples} we make explicit the Katz criterion \citep{katz1978number,katz1979number} for thermodynamic-stability used throughout this paper, while showing how to calculate the life-time of metastable states following \citet{2005A&A...432..117C}. Finally, we reproduce known stability analysis given in the literature both in the canonical and microcanonical ensembles for rather small $R$ values in GR, and show the existence of ensemble inequivalence.
%In \cref{sec:appendix:halo-formation}, we use a Press-Schechter based formalism to obtain the virial mass and radius in a WDM cosmology, for which we previously calculate the matter power spectrum for $mc^2 \sim \SI{10}{\kilo\eV}$ using the CLASS code \citep{2011JCAP...09..032L}. This a necessary step to provide the boundary halo conditions at virialization (starting at $z \sim 10$ in this setup, all the way until $z\sim 2$) to be applied in a self-consistent manner in the thermodynamic-stability studies of \cref{sec:thermo-stability-analysis}. In addition we plot the corresponding virial mass vs. redshift ($z$), of key importance regarding the open issues of DM halo formation and super massive BH formation in the high z Universe.

\section{Self-gravitating fermions at finite temperature in GR}
\label{sec:equilibrium-eqtns}

In this section we introduce the system of equations for a self-gravitating system of massive, neutral fermions (spin 1/2) in hydrostatic equilibrium within GR. More specifically, we solve the Tolman-Oppenheimer-Volkoff (TOV) equations for a perfect fluid whose equation of state (EoS) takes into account: (i) the relativistic effects of the fermionic particles, (ii) finite temperature effects of the system, and (iii) escape of particle effects (i.e. tidally truncated) at large momentum ($p$) through a cut-off in the Fermi-Dirac DF as shown below,

\begin{equation}
	\label{eq:FD-DF-cutoff}
f(r,\epsilon)=\begin{cases}
\frac{1-{\rm e}^{[\epsilon-\epsilon_c(r)]/k_B T(r)}}{{\rm e}^{[\epsilon-\mu(r)]/k_B T(r)}+1} & \text{if $\epsilon<\epsilon_c$},\\
0 & \text{if $\epsilon>\epsilon_c$}
\end{cases}
\end{equation}

where $\epsilon=\sqrt{c^2 p^2+m^2 c^4}-mc^2$ is the particle kinetic energy, $\epsilon_c$ is the cut-off particle energy above which no more particles are left, $\mu$ is the chemical potential with the particle rest-energy subtracted off, and $T$ is the temperature. It is relevant to emphasize that the parameters, $\mu$, $T$ as well as $\epsilon_c$ are all functions of the radius $r$ of the configuration, corresponding with the fulfilment of the Tolman, Klein (i.e. zero and first law of thermodynamics in GR) and energy conservation along a geodetic (see below). We denote by $k_B$ the Boltzmann constant, $h$ is the Planck constant, $c$ is the speed of light, and $m$ is the fermion mass. We do not include the presence of anti-fermions, i.e.~we consider temperatures $T \ll m c^2/k_B$. The full set of ($r$-functional) parameters of the model are defined by the temperature, degeneracy and cutoff parameters, $\beta=k_B T/(m c^2)$, $\theta=\mu/(k_B T)$ and $W=\epsilon_c/(k_B T)$, respectively.

Importantly, the quantum phase-space function \cref{eq:FD-DF-cutoff} can be obtained from a maximum entropy production principle as first shown for fermions in \citep{1998MNRAS.300..981C}. Indeed, it is there shown to be a stationary solution of a generalized Fokker-Planck equation for fermions including the physics of violent relaxation and evaporation, appropriate within the non-linear stages of structure formation. As explained in \cref{sec:introduction}, these results were first conceptualized in the pioneer work of D. Lynden-Bell for collisionless stellar systems \citep{1967MNRAS.136..101L}, and formalized and generalized much later for quantum (collisionless) particles \citep{1996ApJ...466L...1K,1998MNRAS.300..981C,2004PhyA..332...89C}. Indeed, maximum entropy principles applied to DM halo formation in terms of self-gravitating systems, are being recently re-considered in the literature, as in the case of dwarf galaxies in \citep{2020A&A...642L..14S}. 

We write below all the relevant equations to this problem, and leave for next section the thermodynamic (complementary) formalism. The matter source for the corresponding Einstein equations are given in terms of the following (parametric) fermionic EoS
\begin{align}
	\label{rhoepdefscutoff}
	\rho(r)	&= m\frac{2}{h^3}\int_{0}^{\epsilon_c}f_c(r,p)\left(1+\frac{\epsilon(p)}{mc^2}\right){\rm d}^3p\ ,\\
	P(r)	&= \frac{4}{3h^3}\int_{0}^{\epsilon_c}f_c(r,p)\,\epsilon\,\frac{1+\epsilon(p)/2mc^2}{1+\epsilon(p)/mc^2}{\rm d}^3p\ ,
\end{align}
where the integration is carried out over the momentum space bounded from above by the escape energy $\epsilon\leq\epsilon_c$.
%
%\begin{equation}
%	\label{eq:FD-DF-cutoff}
%	f_c(r,\epsilon\leq\epsilon_c) = \ddfrac{1-{\rm e}^{[\epsilon-\epsilon_c(r)]/k_B T(r)}}{{\rm e}^{[\epsilon-\mu(r)]/k_B T(r)}+1}, \qquad
%	f_c(r,\epsilon>\epsilon_c) = 0\,,
%\end{equation}
%

We consider the system as spherically symmetric so we adopt the metric
\begin{equation}
	\label{eq:metric}
	{\rm d}s^2 = {\rm e}^{\nu}c^2 {\rm d}t^2 -{\rm e}^{\lambda}{\rm d}r^2 -r^2 {\rm d}\Theta^2 -r^2\sin^2\Theta {\rm d}\phi^2,
\end{equation}
where ($r$, $\Theta$, $\phi$) are the spherical coordinates, and $\nu$ and $\lambda$ depend only on the radial coordinate $r$.

The Tolman and Klein conditions are:
\begin{align}
	\label{eq:tolman}
	{\rm e}^{\nu/2} T &= {\rm constant},\\
	\label{eq:klein}
	{\rm e}^{\nu/2}(\mu+m c^2) &= {\rm constant}.
\end{align}

The cutoff condition comes from the energy conservation along a geodesic,
\begin{equation}
	\label{eq:energy-conservation}
	{\rm e}^{\nu/2}(\epsilon+m c^2) = {\rm constant},
\end{equation}
that leads to the cutoff (or \textit{escape} energy) condition $(1+W \beta)={\rm e}^{(\nu_b-\nu)/2}$, where $\nu_b\equiv\nu(r_b)$ is the metric function at the boundary of the configuration, i.e. $W(r_b)=\epsilon_c(r_b)=0$, and $r_b\equiv R$ is the boundary radius often called \textit{tidal} radius. The above cutoff formula reduces to the known escape velocity condition ${v_e}^2=-2\Phi$ in the classical limit $c\to\infty$ (i.e. ${\rm e}^{\nu/2}\approx 1+\Phi/c^2$) considered by \citet{1966AJ.....71...64K}, where $V=m\Phi$ with $\Phi$ the Newtonian gravitational potential, adopting the choice $V(r_b)=0$.

The above conditions together with the Einstein equations lead to the system of dimensionless equilibrium equations
\begin{align}
	\label{eq:diff-mass}
	\frac{{\rm d}\hat M}{{\rm d}\hat r}	&= 4\pi\hat r^2\hat\rho(\hat r),\\
	\label{eq:diff-nu}
	\frac{{\rm d}\nu}{{\rm d}\hat r}	&= \frac{2(\hat M(\hat r)+4\pi\hat P\hat r^3)}{\hat r^2(1-2\hat M(\hat r)/\hat r)}, \\
	\label{eq:diff-theta}
	\frac{{\rm d}\theta}{{\rm d}\hat r}	&= -\frac{1-\beta_0(\theta(\hat r)-\theta_0)}{\beta_0}
											   \frac{\hat M(\hat r)+4\pi\hat P(\hat r)\hat r^3}{\hat r^2(1-2\hat M(\hat r)/\hat r)},\\
	\label{eq:beta}
    \beta(\hat r)								&= \beta_0 {\rm e}^{\frac{\nu_0-\nu(\hat r)}{2}}, \\
	\label{eq:cutoff}
    W(\hat r)									&= W_0+\theta(\hat r)-\theta_0\,.
\end{align}
The first two correspond to the only relevant Einstein equations (mass and TOV equations), while the third is a convenient combination of Klein and Tolman relations for the gradient of $\theta=\mu/(k_B T)$. The fourth equation reads for Tolman, and $W(r)$ is a direct combination of Klein and cutoff energy condition of key relevance for thermodynamics (see \cref{sec:cutoff-case}). We have introduced the dimensionless quantities: $\hat r=r/\chi$, $\hat M=G M/(c^2\chi)$, $\hat\rho=G \chi^2\rho/c^2$, $\hat P=G \chi^2 P/c^4$, where $\chi=(\hbar/mc)(m_p/m)$ and $m_p=\sqrt{\hbar c/G}$ the Planck mass. We note that the constants of the Tolman and Klein conditions are evaluated at the center $r=0$, indicated with a subscript \enquote{0}.

The above coupled system of ordinary (highly non-linear) differential \cref{eq:diff-mass,eq:diff-nu,eq:diff-theta,eq:beta,eq:cutoff} implies a boundary condition problem to be solved numerically. It was first solved in \citet{2018PDU....21...82A} for regular conditions at the center of the configuration [$M(0)=0$, $\theta(0)=\theta_0$, $\beta(0)=\beta_0$, $\nu(0)=\nu_0$, $W(0)=W_0$], for given DM particle mass $m$, to find solutions consistent with the DM halo observables of the Galaxy. For positive central degeneracy parameter $\theta_0 > 0$, the DM profiles (fulfilling with fixed halo boundary conditions inferred from observables), develop a \textit{dense~core--diluted~halo} morphology where the central core is sensitive to the particle mass \citep{2015MNRAS.451..622R,2018PDU....21...82A}. While the central core is governed by Fermi-degeneracy pressure, the outer halo holds against gravity by thermal pressure and resembles the King profile or Burkert profile (see \citealp{2018PDU....21...82A,2019PDU....24..278A} for details).

A model built upon the above considerations is usually called as the Ruffini-Argüelles-Rueda (RAR) model, in honour to the authors in \citet{2015MNRAS.451..622R}\,\footnote{
	The underlying system of equations \cref{eq:diff-mass,eq:diff-nu,eq:diff-theta,eq:beta,eq:cutoff} of the original RAR model (i.e. without the cutoff at large momentum $W \to \infty$) was first derived in \citet{1990A&A...235....1G}, though the boundary condition problem relevant for galactic observables was first properly solved in \citet{2015MNRAS.451..622R}.
}, and then extended in \citet{2018PDU....21...82A} including for escape of particles (or tidal effects). Importantly, such a DM halo model is the more general of its kind, given it does not work in the fully Fermi-degeneracy regime $\theta_0 \gg 1$ (as in \citealp{2017MNRAS.467.1515R}), nor in the diluted-Fermi regime $\theta_0 \ll -1$ (as in \citealp{2014MNRAS.442.2717D}). However, in the next sections we will cover all regimes and check along the full set of equilibrium configurations which are thermodynamically stable or unstable, and analyze if they are of astrophysical interest regarding the DM halo phenomenology.

The main advantages of fermionic DM profiles with a \textit{core-halo} morphology (e.g. as in \citealp{2018PDU....21...82A}) over the diluted ones (e.g. Boltzmannian-like), can be summarized as follows:
\begin{enumerate}
    \item the arising of the fermion-degeneracy pressure developed through the center of the DM halo is able to stop the gravitational core-collapse to a singularity, thus preventing the gravothermal catastrophe typical of Boltzmannian phase-space DF. Such a result was first demonstrated in \citet{1998MNRAS.296..569C} in Newtonian gravity, and further shown here as well as in \citet{2020EPJB...93..208A,2019arXiv190810303C} in full GR;
    \item  thermodynamically meta-stable (i.e. local maxima of entropy) \textit{core-halo} solutions of self-gravitating fermions are extremely long-lived, and shown to be of astrophysical relevance when applied to DM halos as demonstrated in this work. Besides, they are more likely to arise in Nature than global entropy maxima (King-like) solutions \citep{2005A&A...432..117C};
    \item  such a \textit{core-halo} DM distribution is in good agreement with overall rotation curve data from dwarf to elliptical galaxies, while the dense core can be an alternative for the central massive BH scenario \citet{2019PDU....24..278A}, including the case of our own Galaxy \citet{2018PDU....21...82A,2020A&A...641A..34B} and for the same particle mass in the keV regime.
\end{enumerate}

In the limit $W \to \infty$ (i.e. $\epsilon_c \to \infty$) the system of equations above reduce to the equations considered in the original RAR model \citep{2015MNRAS.451..622R}. Such a limit, clearly implies that no escape of particles at all is present in the Fermi-Dirac DF (\ref{eq:FD-DF-cutoff}), which is reduced to the typical formula below (where the upper bound integration limit for $p$ in the EoS is set to infinity).
\begin{equation}
	\label{eq:FD-DF}
	f(r,\epsilon) = \ddfrac{1}{{\rm e}^{[\epsilon-\mu(r)]/k_B T(r)}+1},
\end{equation}
The main difference between a \textit{core-halo} DM profile resulting from equations \cref{eq:diff-mass,eq:diff-nu,eq:diff-theta,eq:beta,eq:cutoff} which is built either upon (\ref{eq:FD-DF-cutoff}) or (\ref{eq:FD-DF}), is that in the first case the outer halo resembles a King-like profile, while in the second case it goes as $\rho\propto r^{-2}$ as $r\to\infty$ (resembling a pseudo isothermal-sphere).

\section{Thermodynamic stability analysis of DM halos at virialization}
\label{sec:thermo-stability-analysis}
In the former section we set the necessary equations to obtain a system of self-gravitating fermions in hydrostatic equilibrium, which can be successfully applied to model DM halos as in the RAR model. However, such stability equations do not explain how this kind of collisionless systems reach the steady state in a given cosmological setup. Indeed, the relaxation of collisionless self-gravitating particles, represents a rather complicated problem which involves complex processes such as violent relaxation and nonlinear Landau damping \citep{2008gady.book.....B}. As first shown in the seminal paper of D. Lynden-Bell \citep{1967MNRAS.136..101L}, it is possible to obtain a qSS resulting from a violent relaxation process, by solving a maximization (coarse-grained) entropy problem at fixed total energy and particle number. The concept of thermodynamical stability, at difference to dynamical stability, is thus understood in terms of such a maximization problem. Moreover, it can be shown that if an equilibrium state in GR is thermodynamically stable (i.e. coarse-grained entropy maxima) then is always dynamically stable \citet{1980ApJ...238.1101I}, while in general, the reciprocal is incorrect (see \citealp{2019arXiv190810806C} for a detailed discussion in Newtonian gravity as well as in GR.)

Historically, the main motivation to study the stability of dense clusters composed by self-gravitating objects was linked to the appealing idea that the formation of massive BHs at the center of active galactic nuclei (AGN) could be the the result of the collapse of such clusters \citep{1984ARA&A..22..471R}. The bulk of the works in the field were aimed to the analysis of the relativistic instability (i.e. collapse) of dense \textit{stellar clusters} following Maxwellian energy distributions with specific cut-offs in energy \textit{a la} Zeldovich-Podurets (see e.g. \citealp{1998ApJ...500..217B} and references therein). Such stability analysis were pursued by approximate methods such as the fractional binding energy criterion or the search for maxima in a $M(\rho_0)$ curve, with $\rho_0$ the central density and $M$ the total mass of the configuration. 

In this work we reconsider the problem of collapse of dense and relativistic clusters with application to massive BH formation, though the cluster is now a degenerate compact core made of neutral keV DM fermions, surrounded by a diluted halo composed of the very same particles. Such DM fermions follow a much richer energy distribution function such as the generalized Fermi-Dirac DF in \cref{eq:FD-DF-cutoff}, including for central degeneracy, generic cutoff energy and (effective) temperature free parameters; and which is well motivated since it arises from a maximum entropy production principle (see \cref{sec:equilibrium-eqtns}). We use a rigorous approach to analyze the thermodynamic (and dynamical) stability of the fermionic distributions as the one developed by Katz in \citet{katz1978number} (see \cref{sec:appendix:stability_examples}), which was not applied in \citet{1998ApJ...500..217B} nor in similar contemporary works. Such an analysis is properly compared and connected in \cref{sec:TurningPoint} with the turning point criterion in terms of the traditional $M(\rho_0)$ curve. Even if the stability results presented here show some similitude with respect to the ones done in the past for Maxwellian DFs, they cannot be compared on an equal footing. That is, the more general quantum DF used here imply a double spiraling shape in the caloric curves as in \cref{Mk24}, one of gravothermal catastrophic nature including for degeneracy effects (not present in the classical DF), and the other of relativistic nature.

As mentioned above, we will use throughout this work the Katz criterion (detailed in \cref{sec:appendix:stability_examples}) to find the full set of thermodynamical-stable solutions along the series of (hydrostatic) equilibrium. This is a powerful and rigorous method which relies only in the derivatives of specific caloric curves (e.g. total energy vs. $1/T_\infty$), without the need to explicitly calculate the (rather involving) second order variations in entropy.\footnote{
	Interestingly, by just solving the extremization of entropy at fixed energy and particle number (i.e. solving up to first order variation in $S$) it is possible to obtain the Fermi-Dirac DF at statistical equilibrium --- with or without cutoff as used here --- as well as the GR EoS \citep{2019arXiv190810806C}.
} Indeed, this is a commonly used method in the context of self-gravitating systems as can be seen in recent works \citep{2015PhRvD..91f3531C,2020EPJB...93..208A}. In next subsections we work in the microcanonical ensemble, applied for isolated systems so that its energy $E$ is conserved. In this ensemble the relevant thermodynamic potential to be extremized is the coarse-grained entropy $S$.\footnote{
	In \cref{sec:appendix:stability_examples} we complement the analysis in the canonical ensemble for other $N$, $R$ examples, where the relevant thermodynamic condition is the minimization of free energy $F$ at fixed $N$ and $T_\infty$.
} This choice is justified, being the microcanonical ensemble the more appropriate for astrophysical applications as carefully explained in \citet{2015PhRvD..92l3527C}.

Self-gravitating solutions with given particle number $N$ and total energy $E=c^2M(R)$ which extremize the coarse-grained entropy, must be bounded in radius. Otherwise, as the total mass of the system has no upper bound, the maximization entropy problem is not well defined and the entropy will never reach a maximum \citep{2008gady.book.....B}.

In this section we formally perform two thermodynamic stability analysis, under two different choices of the fermionic phase space DFs. In \cref{sec:box-case} we assume a DF given by \cref{eq:FD-DF}. Under this choice, the solutions obtained from \cref{eq:diff-mass,eq:diff-nu,eq:diff-theta,eq:beta,eq:cutoff} have no spatial boundaries, implying DM density profiles scaling as $\rho\propto r^{-2}$ at large distances (see \cref{perf_a15} and e.g. \citealp{2015MNRAS.451..622R} for more details). Thus, as it is customary, we confine such a self-gravitating system within a spherical-box of total radius $R$ in order to avoid an entropy runaway.

In \cref{sec:cutoff-case} instead, we assume a DF given by \cref{eq:FD-DF-cutoff}, corresponding to the more realistic tidally truncated self-gravitating system, naturally bounded in radius ($R\equiv r_b$) thanks to the particle-energy cutoff condition.

Next we introduce the basic thermodynamic potentials in a GR framework, relevant to develop the corresponding stability analysis for the case of self-gravitating fermions at finite $T_\infty$. When working in a curved space-time, the relevant thermodynamic potential is the Gibbons-Hawking free energy \citep{1977PhRvD..15.2752G}
\begin{equation}
	\label{eq:FGH}
	F = M(R) + \int_\Sigma s T u^\alpha {\rm d}\Sigma_\alpha,
\end{equation}
where $M(R)$ is the total mass of the system as obtained by integrating \cref{eq:diff-mass} up to $R$, $\Sigma$ is the spacelike hypersurface occupied by the system (within a sphere of radius $R$) with $u^\alpha$ a unit timelike normal vector, and $s$ is the entropy density which in the case of a relativistic fluid is given by the Gibbs-Duhem relation
\begin{equation}
    \label{eq:ss}
    s(r)=\frac{P(r)+\rho(r)-\mu(r) n(r)}{T(r)}.
\end{equation}
Taking into account the metric tensor (\ref{eq:metric}), and the Tolman and Klein equations, the free energy of the system can be written as a dimensionless expression explicitly dependent on the free parameters of the model,
\begin{equation}
	\label{eq:Fp}
	\begin{split}
		F = &\hat M(\hat R)+\hat m[1+\theta_0\beta_0] {\rm e}^{\nu_0/2}\hat N - \\
		    &\int_{0}^{\hat R} 4\pi \hat r^2 {\rm e}^{(\nu(\hat r)+\lambda(\hat r))/2}[\hat P(\hat r)+\hat\rho(\hat r)] {\rm d}\hat r,
	\end{split}
\end{equation}
where ${\rm e}^{\lambda(\hat r)}= (1-2\hat M(\hat r)/\hat r)^{-1}$ is the space-like metric factor. Similarly, the dimensionless entropy reads
\begin{equation}
	\label{eq:Se}
	\begin{split}
		S = &-\left(\frac{1}{\beta_0}+\theta_0\right)\hat N+ \\
		    &\frac{{\rm e}^{-\nu_0/2}}{\hat m\beta_0}\int_{0}^{\hat R} 4\pi \hat r^2 {\rm e}^{(\nu(\hat r)+\lambda(\hat r))/2}[\hat P(\hat r)+\hat\rho(\hat r)] {\rm d}\hat r.
	\end{split}
\end{equation}
Note that \cref{eq:Fp} can be written as $F=\hat M(\hat R)-\hat T_\infty S$ which is a more familiar expression, reminiscent of classical thermodynamics, where $\hat T_\infty\equiv \beta_\infty$ represents \textit{the} (dimensionless) temperature of the system seen by an observer at infinity.

\subsection{Box-confined DM halos}
\label{sec:box-case}
As explained above, when making a thermodynamical stability analysis with the standard Fermi-Dirac DF (\ref{eq:FD-DF}), it is mandatory to bound the system in a box of radius $R$ in order for $S$ to reach a maximum. Thus, the physical parameters needed to be fixed along the series of equilibrium solutions (extremum of $S$), are in this case the total particle number $N$, and spherical box radius $R$. That is, two extra equations are needed to be added to the system described by \cref{eq:diff-mass,eq:diff-nu,eq:diff-theta,eq:beta,eq:cutoff}
\begin{align}
	\label{eq:N}
	\int_{0}^{\hat R} 4\pi \hat r^2 {\rm e}^{\lambda(\hat r)/2}\hat n(\hat r) {\rm d}\hat r = \hat N,\\
	\label{eq:Schw}
	{\rm e}^{\nu(\hat R)} = 1-\frac{2\hat M(\hat R)}{\hat R}.
\end{align}
Both are given in dimensionless form (with $\hat N = N (m/m_p)^3$), and the second equation being the Schwarzschild condition assuring the continuity of the metric at the boundary radius. The parameters ($\hat N$, $\hat R$) are chosen so to fulfil with the virial mass and radius of a DM halo at virialization in a realistic cosmological setup. Indeed, such parameters are obtained from a Press-Schechter based formalism within a WDM cosmology for $mc^2=\SI{10}{\kilo\eV}$ as detailed in \cref{sec:appendix:halo-formation}. We provide two relevant examples (we shall use $\hat R \equiv \hat R_{\rm vir}$ from now on): \begin{enumerate}[label=\textbf{(\arabic*)}]
	\item $\hat N=0.38, \hat R_{\rm vir}=\num{1.4E7}$ \label{item:example1}
	\item $\hat N=3.67, \hat R_{\rm vir}=\num{3.8E7}$ \label{item:example2}
\end{enumerate} Such values imply, in dimensionfull units, the following virial mass and radius: a typically small DM halo with $M_{\rm vir} \equiv M(R_{\rm vir}) \approx \SI{6.2E9}{\Msun}$ and $R_{\rm vir} = \SI{11.1}{\kilo\parsec}$ corresponding to the example \ref{item:example1}; and an average-sized DM halo with $M_{\rm vir} \equiv M(R_{\rm vir}) \approx \SI{5.9E10}{\Msun}$ and $R_{\rm vir} = \SI{29.7}{\kilo\parsec}$ for \ref{item:example2}. That is, each pair ($R_{\rm vir}$, $M_{\rm vir}$) is consistent with the values obtained from the Press-Schechter formalism as required (see \cref{fig:SecPS_MR200_MFS} in \cref{sec:appendix:halo-formation}). 

The values for $\hat N$ are chosen this way in order to have one case, i.e. \cref{item:example1}, with $\hat N=0.38 < \hat N_{\rm OV}$, and the other, i.e. \cref{item:example2}, with $\hat N=3.67 > \hat N_{\rm OV}$, being $\hat N_{\rm OV} \approx 0.4$ the Oppenheimer-Volkoff critical particle number \citet{1939PhRv...55..374O}. Such a critical value triggering core-collapse, can be written in a more familiar way in terms of the Planck mass $m_{p}$ and the fermion mass $m$ as $N_{\rm OV}=0.398\,m_{p}^3/m^3$, corresponding to a critical mass $M_{\rm OV}=0.384\,m_{p}^3/m^2$ which for $mc^2=\SI{10}{\kilo\eV}$ reads $M_{\rm OV}\approx \SI{6E9}{\Msun}$. Thus, any equilibrium configuration of fermions at finite $T_\infty$ with $M \gtrsim \SI{6E9}{\Msun}$, may undergo (under certain conditions) a core-collapse towards a SMBH as explained in \cref{sec:TurningPoint}. The astrophysical and cosmological consequences of such a core-collapse in terms of the typical DM halo examples considered here, are further commented in \cref{sec:TurningPoint} and \cref{sec:conclusion}.

We next apply the Katz criterion (see \cref{sec:appendix:stability_examples} for the details) to find all the thermodynamically stable branches of solutions along the microcanonical caloric curve $1/\hat T_\infty$ vs. $\hat E$, with  $1/\hat T_\infty=\partial S/\partial \hat E$ and $\hat E=\hat M(\hat R)$.\footnote{
	We notice that in order to recover the non-relativistic limit of the caloric curve, one should re-define the energy as the binding energy $E_b=(M-mN)c^2$, though all the results of this paper hold since the behaviour of the caloric curves and the sign-change of its derivatives around the turning points, are not altered by adding a constant term to the energy.
} Thermodynamically-stable solutions are maxima of entropy $S$ (either local or global) at fixed $E$ and $N$, while the unstable solutions are either minimum or saddle-point of entropy. Importantly, thermodynamically-stable solutions (associated with second order variations of $S$), are only a reduced subset along the full set of (hydrostatic) equilibrium solutions. Indeed, the latter corresponds with solutions arising from the system given \cref{eq:diff-mass,eq:diff-nu,eq:diff-theta,eq:beta,eq:cutoff,eq:N,eq:Schw}, which are simply an extremum of entropy (i.e. to first order $\delta S=0$ for given $N$ and $E$) as demonstrated in \citet{2019arXiv190810806C}. This last statement clearly justifies the need to apply the Katz criterion to make a stability analysis.

%\footnote{
%	More generally, it can be shown that the TOV equation together with the Klein and Tolman relations of \cref{sec:equilibrium-eqtns} can be obtained from such an extremization entropy problem \citep{2019arXiv190810806C}.}

\subsubsection{$M_{\rm vir} \approx \SI{6.2E9}{\Msun}$, $R_{\rm vir} = \SI{11.1}{\kilo\parsec}$}
\label{Box1}
The numerical problem consists in solving \cref{eq:diff-mass,eq:diff-nu,eq:diff-theta,eq:beta,eq:cutoff} under the choice of the DF given by \cref{eq:FD-DF}, with the specific boundary conditions \cref{eq:N,eq:Schw} at virialization. As a first example we consider a rather small halo with $M_{\rm vir} \approx \SI{6.2E9}{\Msun}$, $R_{\rm vir} = \SI{11.1}{\kilo\parsec}$. We solve this problem for a wide range of control parameters ($\nu_0$, $\beta_0$, $\theta_0$), for fixed $mc^2=\SI{10}{\kilo\eV}$, and plot in \cref{fig:calcurvebox1} all the equilibrium solutions (i.e. extremum of $S$) along the caloric curve (i.e. $-\hat M$ vs. $1/\hat T_\infty$) as customary. This problem implies a monoparametric family of solutions, since for a fixed particle mass $m$ we have 3 free model parameters and two given boundary conditions given by \cref{eq:N,eq:Schw}

We differentiate in \cref{fig:calcurvebox1} among the full family-set of thermodynamically-stable solutions (in continuous-blue line), from the thermodynamically-unstable ones which are shown in dotted-violet. We then analyze in detail all the different kind of density profiles (extremum of $S$) for a fixed value of the total energy $\hat M$ inside the spiral structure (see vertical dashed line in \cref{fig:calcurvebox1}). The main conclusions out of this stability analysis can be summarized as follows:
\begin{figure}
	\centering
	\includegraphics[width=\columnwidth, clip]{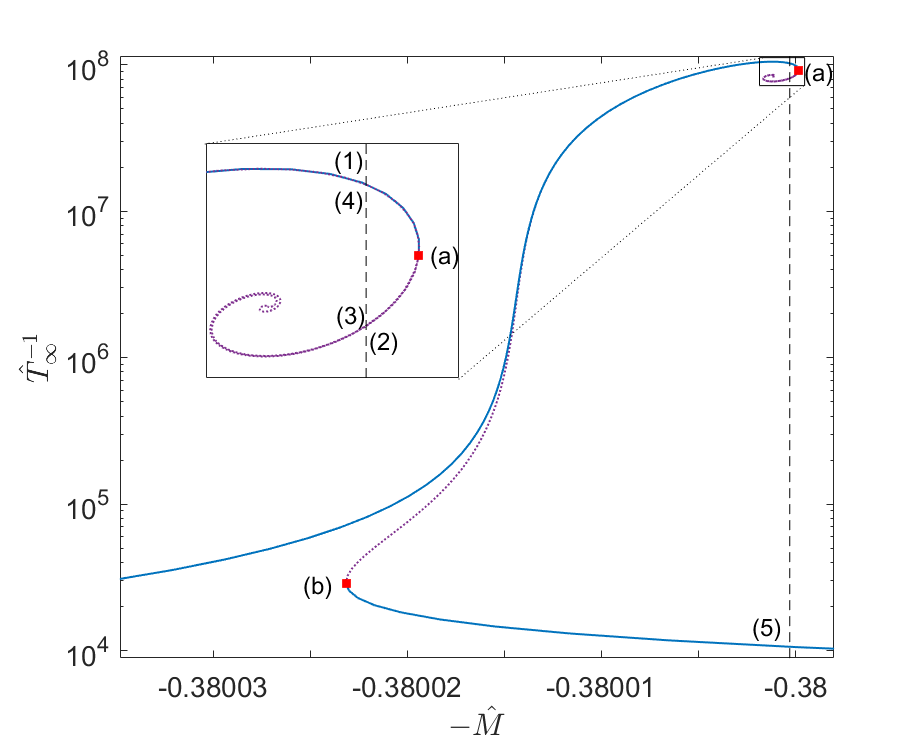}
	\caption{Series of equilibrium solutions along the caloric curve for box-confined configurations of $mc^2=\SI{10}{\kilo\eV}$ fermions fulfilling the boundary conditions \ref{item:example1} at halo virialization. For this large value of $\hat R\sim 10^7$ the curve spirals inwards several turns until it starts unwinding just when the quantum degeneracy (i.e. Pauli principle) sets in at $\theta_0 \gtrsim 10$. The states within the continuous-blue branches are thermodynamically (and dynamically) stable (i.e. either local or global entropy maxima), while the dotted-purple branch - between (a) and (b)- is unstable (i.e. either minimum or saddle point of entropy), according to the stability criterion of \cref{sec:appendix:stability_examples}.} 
	\label{fig:calcurvebox1}
\end{figure}
\begin{figure}
	\centering
	\includegraphics[width=\columnwidth, clip]{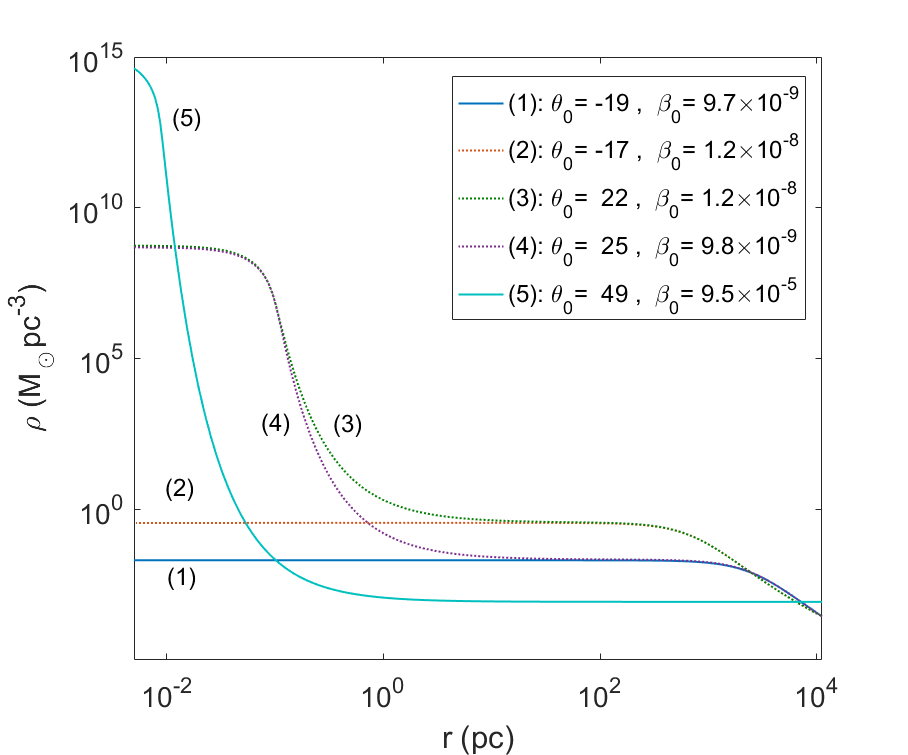}
	\caption{Density profiles for $mc^2=\SI{10}{\kilo\eV}$ corresponding with the equilibrium states of the caloric curve in \cref{fig:calcurvebox1} with energy $\hat M\approx 0.38$. Only the profile (1) (resembling a pseudo-isothermal sphere) and the \textit{core-halo} one (5) are stable, while profiles (2)--(4) are thermodynamically unstable. Solutions like (4) were applied to describe DM halos in galaxies in \citet{2015MNRAS.451..622R}, though are very unlikely to occur in Nature (i.e. unstable), at difference with the more realistic \textit{core-halo} solutions analyzed in \cref{sec:cutoff-case}.}
	\label{perf_a15}
\end{figure}
\begin{enumerate}
    \item Upper continuous-blue branch contains solutions which are entropy maxima (either local or global, see \cref{sec:appendix:stability_examples}), until the point where the caloric curve starts to rotate clockwise (i.e. first point where the curve $1/\hat T_{\infty}$ vs. $-\hat M$ is tangent to a vertical line), labeled as a red-square (a). Interestingly, all the solutions lying in this stable branch belong to the diluted-Fermi regime, with $\theta_0 \ll -1$ (i.e. Boltzmannian-like). The associated density profiles resemble pseudo-isothermal spheres, with a density tail behaviour of $\rho\propto r^{-2}$ at large distances (see curve (1) in \cref{perf_a15});
    \item From this point all along the dotted-violet branch ending in point (b), solutions are thermodynamically-unstable (either minimum or saddle-point of $S$). Hydrostatic equilibrium solutions well inside the spiral start to transition from diluted-Fermi (see e.g. the density profile (2) in \cref{perf_a15} with $\theta_0 \lesssim -1$) to semi-degenerate Fermi (see e.g. density profile (3) with $\theta_0 \gtrsim 10$), just when the caloric curve starts to rotate anticlockwise. Precisely at this point the central core becomes degeneracy pressure supported, that is, the thermal de Broglie wavelength ($\lambda_B$) is larger than the interparticle mean distance at the core $\lambda_B > l_c$. This unwinding within the upper spiral of the caloric curve proves that the Pauli exclusion principle halts the classical gravothermal catastrophe, as first realized within Newtonian gravity in \citet{1998MNRAS.296..569C}. In the \enquote{unrolled} side of the curve, the profiles develop a \textit{core-halo} behaviour, such that the larger $\theta_0\gtrsim 10$ the more compact and degenerate is the central core while the more diluted and extended the outer halo. Unstable \textit{core-halo} solutions like (4) are similar to the ones applied to DM halos in \citet{2015MNRAS.451..622R} within the original RAR model (see \cref{sec:comparison} for further relevant implications);
    \item For larger energies (at the right of point (b), where the amount of anticlockwise rotations of the caloric curve equals the number of clockwise rotations), a lower branch of entropy maximum (either local or global) arises, labelled in continuous-blue. The thermodynamically meta-stable solutions of this branch (i.e. local entropy maxima), have a tremendously long-lived life-time much larger than the age of the Universe as calculated in \cref{sec:appendix:stability_examples}. The interesting connection between meta-stability and long-liveness of self-gravitating systems with long range interactions was originally demonstrated in \citet{2005A&A...432..117C}. Typical stable density profiles (see e.g. density profile (5) of \cref{perf_a15}) develop highly compact core sizes below \si{\milli\parsec} scales, surrounded by a very extended and diluted halo with no relevant astrophysical application as further discussed below;
    \item We recognize the existence of a spiral-feature in the caloric curve, with the absence of the monotonic inspiraling typical of Boltzmannian DF (see e.g. \citealp{1990PhR...188..285P}). This necessarily implies that the gravothermal catastrophe is avoided, since the singular isothermal-sphere is never present along our fermionic family of solutions. Such a result was first demonstrated in \citet{1998MNRAS.296..569C} for the Fermi-Dirac DF in Newtonian gravity and generalized here within GR for realistic DM halo-sizes (see also \citealp{2020EPJB...93..208A});
    
    \item By comparing the observationally inferred DM surface density $\Sigma_{0{\rm D}}^{\rm obs} \sim \SI{E2}{\Msun\per\parsec^2}$ \citep{2009MNRAS.397.1169D} (including $3-\sigma$ errors in orange band) with the theoretical prediction\footnote{
      \label{fn:donato}
      Such predicted magnitude is calculated as $\Sigma_{0{\rm D}}\propto \rho(r_{p})r_h$, with $\rho(r_{p})\equiv \rho_p$ the density at plateau and $r_h$ the halo scale-length as detailed in \citet{2019PDU....24..278A}).
    } from the corresponding fermionic density profiles along the caloric curve, we conclude: 
    \begin{itemize}
    \item There are no thermodynamically-stable \textit{core-halo} solutions of astrophysical interest when applied to low mass ($\sim \SI{E9}{\Msun}$) DM halos (box-confined). Either they can fit DM halo observables (like the solution (4) as obtained within the RAR model model in \citealp{2015MNRAS.451..622R}) though are saddle points of entropy; or they are (local) maxima of entropy as solution (5), but the halo is too diluted to match the observational $\Sigma_{0{\rm D}}$ relation.
    \item There exist diluted-Fermi DM profiles (like solution (1) resembling pseudo-isothermal spheres) which are stable (i.e. maxima of entropy), and at the same time they match the observational $\Sigma_{0{\rm D}}$ relation. These statements can be directly checked by comparing the information in \cref{fig:calcurvebox1,perf_a15,sigma_a15}.
    \end{itemize}
\end{enumerate}
%
% Thus, for $\hat M \approx 0.38$ (vertical dashed line), only density profile solutions (1) and (4) are of astrophysical interest regarding the DM (inferred) observable $\Sigma_{0{\rm D}}$, but only (1) is thermodynamically-stable, while (4) is unstable (i.e. unreachable).

\begin{figure}
	\centering
	\includegraphics[width=\columnwidth, clip]{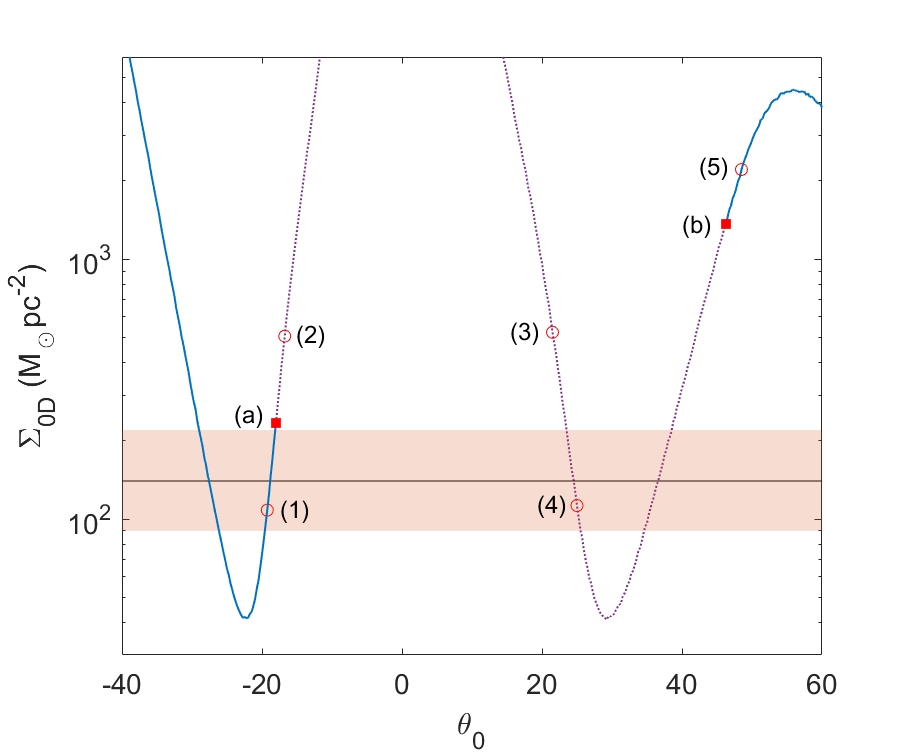}
	\caption{The observationally inferred DM surface density $\Sigma_{0{\rm D}}^{\rm obs} \sim \SI{E2}{\Msun\per\parsec^2}$ (including $3-\sigma$ errors in orange band, \citealp{2009MNRAS.397.1169D}) is compared with the theoretical prediction $\Sigma_{0{\rm D}}\propto \rho_p r_h$ (see footnote \ref{fn:donato}), for the full series of equilibrium along the caloric curve of \cref{fig:calcurvebox1}. Only pseudo-isothermal-like solutions such as (1) (see \cref{perf_a15}) are stable and agree with $\Sigma_{0{\rm D}}^{\rm obs}$ at the same time.}
	\label{sigma_a15}
\end{figure}

\subsubsection{$M_{\rm vir}\approx \SI{5.9E10}{\Msun}$, $R_{\rm vir} = \SI{29.7}{\kilo\parsec}$}
\label{Box2}
\begin{figure}
	\centering
	\includegraphics[width=\columnwidth]{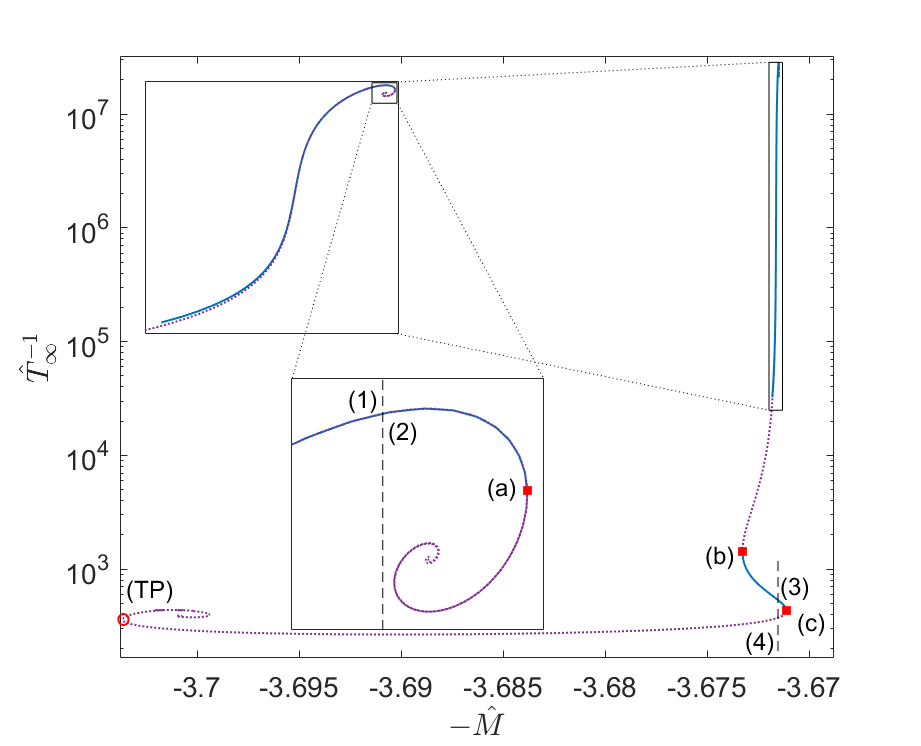}
	\caption{Series of equilibrium solutions along the caloric curve for box-confined configurations of $mc^2=\SI{10}{\kilo\eV}$ fermions fulfilling with the boundary conditions \ref{item:example2} at halo virialization. The states within the continuous-blue branches are thermodynamically (and dynamically) stable (i.e. either local or global entropy maxima), while the dotted-violet branches - between (a) and (b) and after (c) - are unstable (i.e. either minimum or saddle point of entropy), according to the stability criterion explained in \cref{sec:appendix:stability_examples}. The arising of the second spiral of relativistic origin for high $T_\infty$ is characteristic of caloric curves at fixed $N$ within GR, and imply the existence of a turning point in a $M(\rho_0)$ curve (see \cref{sec:TurningPoint})}.
	\label{Ma16}
\end{figure}

\begin{figure}
	\centering
	\includegraphics[width=\columnwidth]{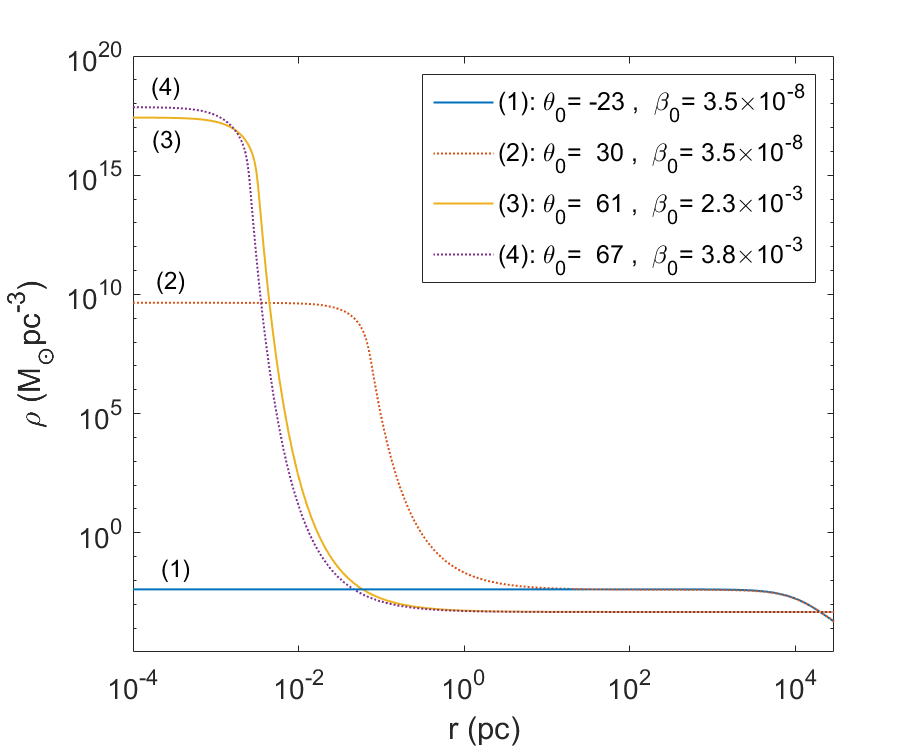}
	\caption{Density profiles for $mc^2=\SI{10}{\kilo\eV}$ corresponding with the equilibrium states of the caloric curve in \cref{Ma16} with energy $\hat M\approx 3.672$. Only the profiles (1) (resembling a pseudo-isothermal sphere) and the \textit{core-halo} one (3) are stable, while profiles (2) and (4) are thermodynamically unstable. Solutions like (2) were applied to describe DM halos in galaxies similar to the Milky Way in \citet{2015MNRAS.451..622R}, though are very unlikely to occur in Nature (i.e. unstable), at difference with the more realistic \textit{core-halo} solutions analyzed in \cref{sec:cutoff-case}.}
	\label{perfa16}
\end{figure}

\begin{figure}
	\centering
	\includegraphics[width=\columnwidth]{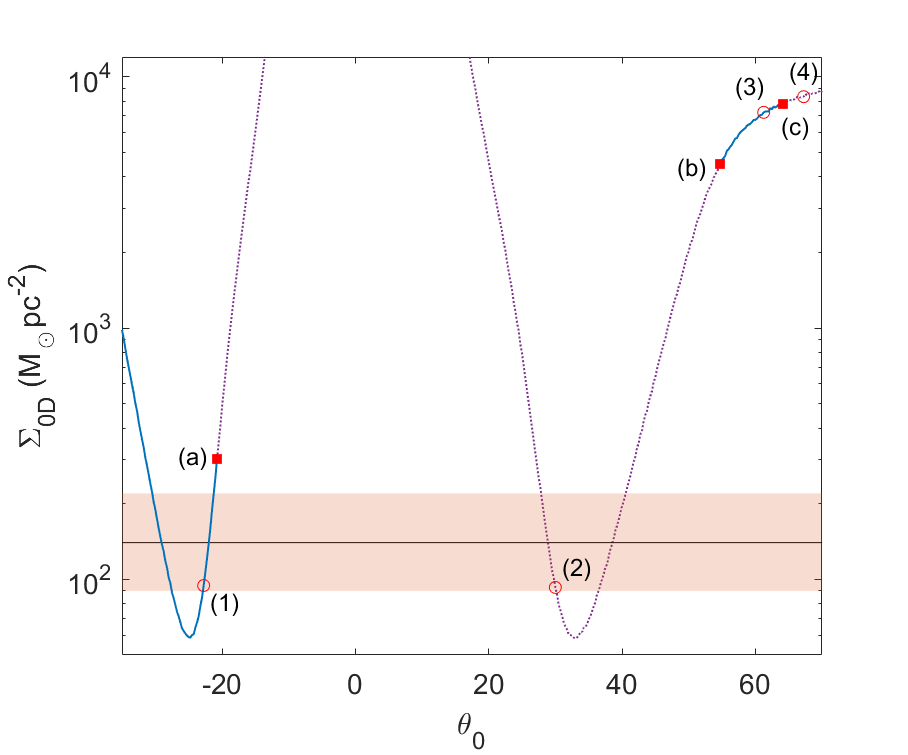}
	\caption{The observationally inferred DM surface density $\Sigma_{0{\rm D}}^{\rm obs} \sim \SI{E2}{\Msun\per\parsec^2}$ (including $3-\sigma$ errors in orange band, \citealp{2009MNRAS.397.1169D}) is compared with the theoretical prediction $\Sigma_{0{\rm D}}\propto \rho_p r_h$ (see footnote \ref{fn:donato}), for the full series of equilibrium along the caloric curve of \cref{Ma16}. Only pseudo-isothermal-like solutions such as (1) (see \cref{perfa16}) are stable and agree with $\Sigma_{0{\rm D}}^{\rm obs}$ at the same time.}
	\label{sigma_a16}
\end{figure}

We repeat here the same analysis done in the above example, but in this case the total particle number correspond to larger halos, exceeding the OV limit (as e.g. $\hat N = 3.67 > \hat N_{\rm OV}$). This implies a novel qualitative behaviour in the lower end of the caloric curve, towards the more degenerate and relativistic (i.e. higher $T_\infty$) configurations, while for lower values of such parameters the situation is similar than in the above $\hat N < \hat N_{\rm OV}$ case. Indeed, by comparing the \cref{Ma16,perfa16,sigma_a16} with respect to the analogous \cref{fig:calcurvebox1,perf_a15,sigma_a15}, it can be directly concluded that the main five results drawn in the above example with $\hat N < \hat N_{\rm OV}$ hold here as well, as we summarize below.

\begin{enumerate}
    \item Starting with stable (entropy maxima) states corresponding to solutions with $\theta_0 \ll -1$, (in upper continuous-blue branch of \cref{Ma16}), there is an equal amount of times the caloric curve rotates clockwise starting at (a) (i.e. gaining an instability mode) as rotates anticlockwise (i.e. recovering a stability mode). This process continues until point (b), where the thermodynamic stability is restored. In between the points (a) and (b) the solutions are either minimum or saddle points of entropy, and progressively increase their central degeneracy until a typical \textit{core-halo} profile (i.e. with $\theta_0 \gtrsim 10$) is obtained just after the spiral unwinds.
    
    \item after point (b) and up to (c), the solutions are entropy maxima (either local or global) and therefore thermodynamically meta-stable or stable respectively. Meta-stable solutions are shown to be extremely long lived as calculated in \cref{sec:appendix:stability_examples}, and thus can be considered as stable and even more likely to occur in Nature than global entropy maxima as explained in \citet{2005A&A...432..117C}. Similarly as in the above case, these stable \textit{core-halo} solutions are of no astrophysical interest when applied to typical DM halo masses ($\sim \SI{E10}{\Msun}$), since their halo densities are too diluted (and their sizes too extended) to match the observational $\Sigma_{0{\rm D}}$ relation (see e.g. solution (3) in \cref{perfa16,sigma_a16}). 
    
    \item at difference with the former case ($\hat N < \hat N_{\rm OV}$), a turn-over occurs in the caloric curve at the ending point of the stable branch where the continuous-blue line is tangent to a vertical line (labelled (c) in \cref{Ma16}). According to the Katz criterion (see \cref{sec:appendix:stability_examples}) a new instability mode is gained because at (c) an extra  clockwise rotation takes place in the thermodynamic curve. Left of this point arises a new branch of thermodynamically-unstable solutions, plotted in \cref{Ma16} in dotted-violet, which extends indefinitely into the second spiral in the caloric curve.\footnote{
        A spiraling of relativistic origin (similar to our case) in a caloric curve was first shown in \citet{2015CQGra..32m5023R} for a self-gravitating ideal gas confined in a spherical box, and extended in \citet{2019CQGra..36f5001R} for fermions at finite $T$ in GR.
    } Importantly, at the bottom-left end of this unstable branch we recognize the so called \enquote{turning point} (TP) instability (labelled with an empty red circle), whose mathematical, physical and astrophysical properties are discussed in \cref{sec:TurningPoint}.
\end{enumerate}


\paragraph*{Remark 1} The pseudo-isothermal sphere-like behaviour (belonging to the diluted-Fermi class of stable configurations), implies density halo tails going as $\rho \sim r^{-2}$ for large $r$, until the abrupt end of the boundary-box radius $R$ sets in. Though such a slope is in tension with standard DM halo phenomenology, which indicates sharper halo trends such as $\sim r^{-3}$ as in the Burkert or NFW profiles. Importantly, halo tails more similar to this phenomenologically suitable ones, can be obtained in the more realistic tidally truncated scenario, while at the same time belonging to stable branch, shown in the following \cref{sec:cutoff-case}.

\paragraph*{Remark 2} As shown in \citet{2020EPJB...93..208A}, for box-confined configurations of fermions, the larger the box size $\hat R$ (as compared to $R_{\rm OV}$), the less relevant are the GR effects in the overall distribution when compared with Newtonian gravity. As we are using very large values of $\hat R\sim 10^7$ in order to reach realistic galactic sizes for $\mathcal{O}(\SI{10}{\kilo\eV})$ particles, very similar conclusions about the stability should be obtained within Newtonian equations as well. However, we notice that appreciable differences should arise towards the high $\hat T_\infty$ end of the caloric curve, since the radius of the degenerate core is close to the critical one before collapse\footnote{Indeed, within the theory of degenerate stars (resembling the compact cores in our configurations), an excess of $\sim 10\%$ in the critical mass at collapse is known to arise if the analysis is done in Newtonian gravity (though with relativistic energies) instead of GR \citep{1974bhgw.book.....R}.} (i.e. notice that $r_c \sim R_{\rm OV}$ close to point (c) of \cref{Ma16}; see also \cref{sec:TurningPoint}).   

\subsection{Tidally truncated DM halos}
\label{sec:cutoff-case}

In tidally truncated systems, the tidal radius $R$ which is naturally set by the escape energy condition in \cref{sec:equilibrium-eqtns}, is \textit{not} fixed along the series of equilibrium. Indeed, it must be sensitive to the free parameters of the system which variate along the caloric curve, including $T_\infty$ . What remains constant instead, is a combination of the Klein and escape energy conditions, as explicited in \cref{sec:equilibrium-eqtns} and \cref{eq:cutoff}. 

In other words, at difference with the box-confined case, the factor $W_0-\theta_0=(\epsilon_c-\mu)/(k_B T)$ is the one to be kept constant here, together with $N$. It is only by fixing this factor (and not $R$) that the Katz criterion applies, and that the changes of stability in the thermodynamic curve (e.g in the microcanoncial ensemble) correspond to the maximization entropy problem \citep{1980MNRAS.190..497K,2015PhRvD..92l3527C}.\footnote{
	This fact was first recognized by Katz \citep{1980MNRAS.190..497K}, and detailed further in \citet{2015PhRvD..92l3527C} who used the dimensionfull constant factor $A=(2 m^4/h^3){\rm e}^{-(\epsilon_c-\mu)/k_B T}$ proportional to the one fixed here.
}

In tidally-truncated systems, the spanning of the caloric curve (either in temperature as in energy) is sensitive to the constant $W_0-\theta_0$, affecting as a consequence the different locations of the solutions in such curve. Given such solutions may be of astrophysical interest for DM halo applications, it is thus convenient to have an educated guess on which $W_0-\theta_0$ value to start with, in order not to make (undesired) blind trials. 

Fortunately, the phenomenology of different observationally inferred DM halos (from dwarf all the way to elliptical galaxies) in terms of a self-gravitating systems of fermions given by the equations \cref{eq:diff-mass,eq:diff-nu,eq:diff-theta,eq:beta,eq:cutoff}, was already worked out in \citet{2019PDU....24..278A} within the RAR model. In that work a particle mass of $mc^2=\SI{48}{\kilo\eV}$ was used as a relevant example, in views of the successful Milky Way phenomenology developed within the same model in \citet{2018PDU....21...82A}. For halos with density tails eventually acquiring a power law $r^{-n}$ with $n>2$ (i.e. close to the ones provided by N-body simulations), we obtain from \citet{2019PDU....24..278A} typical $W_0-\theta_0$ values between $20$ and $28$ (and reaching up to $40$ or larger for isothermal sphere density tails $\propto r^{-2}$). 

Therefore, we consider next the boundary condition problem ($\hat N, W_0-\theta_0$), and perform the corresponding thermodynamic stability analysis (for $mc^2=\SI{48}{\kilo\eV}$ as motivated in the comment above), with the following values: $\hat N=76.25$, $W_0-\theta_0=24$.

The chosen value for $\hat N=76.25 > \hat N_{\rm OV}$) implies, in dimensionfull units, the virial mass of an average DM halo mass $M_{\rm vir} \equiv M(R_{\rm vir}) \approx \SI{5.4E10}{\Msun}$. That value being chosen this way in order to properly compare the thermodynamical stability results of this section with the ones of \cref{Box2}.

With the educated choice of $W_0-\theta_0=24$, it is expected to find along the series of equilibrium, a solution with a halo size at virialization of $R_{\rm vir} \approx \SI{29}{\kilo\parsec}$ (as dictated from \cref{fig:SecPS_MR200_MFS} within the Press-Schechter formalism of \cref{sec:appendix:halo-formation}). Besides that, it is further pursued to check if such an expected solution may be both, thermodynamically stable and of astrophysical interest or otherwise.  

The numerical problem is somewhat analogous as the case \cref{Box2}. Indeed, for the tidally truncated halos under consideration (i.e. with $\hat N > \hat N_{\rm OV}$), it is expected that the caloric curve to be qualitatively similar to the one in the box-confined case (see \citealp{2015PhRvD..92l3527C} for an analogous comparison in Newtonian gravity reaching the same conclusion). However, it is clear that the precise locations of the points where stability changes, as well as the stability-branch extensions along the caloric curve, should shift with respect to one another. It is thus worthy to make here a detailed investigation to search for new family of solutions, with the hope to find profiles which are thermodynamically stable as well as of astrophysical interest (not possible for the box-confined case as shown in \cref{sec:box-case}).

\begin{figure}
	\centering
	\includegraphics[width=\columnwidth]{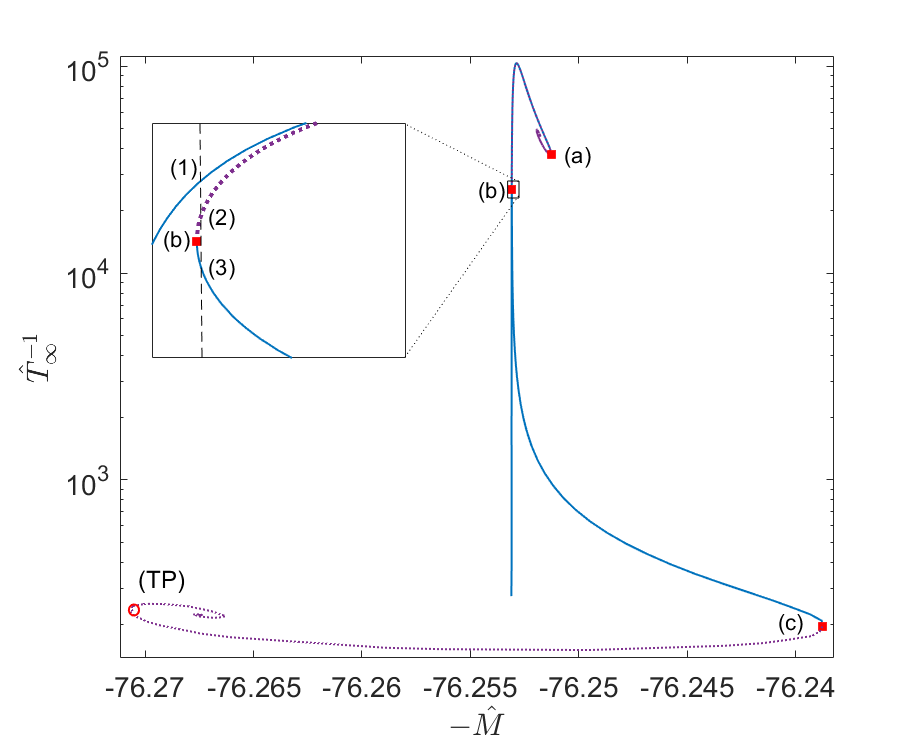}
	\caption{Series of equilibrium solutions along the caloric curve for tidally-truncated configurations of $mc^2=\SI{48}{\kilo\eV}$ fermions fulfilling with ($\hat N=76.25$, $W_0-\theta_0=24$). The states within the continuous-blue branches are thermodynamically (and dynamically) stable (i.e. either local or global entropy maxima), while the dotted-violet branches - between (a) and (b) and after (c) - are unstable (i.e. either minimum or saddle point of entropy), according to the Katz criterion. Solution (3) is stable and fulfills with the virialization conditions as required from the Press-Schechter formalism of \cref{sec:appendix:halo-formation}. The arising of the second spiral of relativistic origin for high $T_\infty$ is characteristic of caloric curves at fixed $N$ within GR, and imply the existence of a turning point in a $M(\rho_0)$ curve (see \cref{sec:TurningPoint})}.
	\label{Mk24}
\end{figure}

\begin{figure}
	\centering
	\includegraphics[width=\columnwidth]{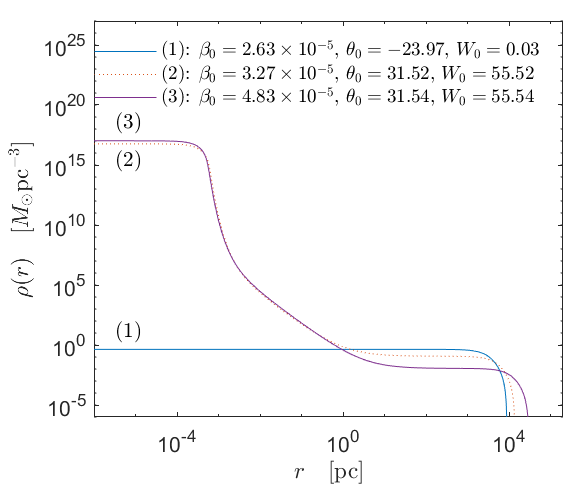}
	\caption{Density profiles for $mc^2=\SI{48}{\kilo\eV}$ corresponding with the equilibrium states of the caloric curve in \cref{Mk24} with energy $\hat M\approx 76.25308$. Only the profiles (1) (resembling a King distribution) and the \textit{core-halo} one (3) are stable, while profile (2) is thermodynamically unstable. Interestingly, solutions like (3) were successfully applied to explain the DM halo in the Milky Way in \citet{2018PDU....21...82A}. They are stable, extremely long-lived and fulfil the $\Sigma_{0{\rm D}}^{\rm obs}$ relation as shown in \cref{Sigma0Dk24}, as well as the expected value of the dispersion velocity $\sigma_h$ in CDM N-body simulations as shown in \cref{Sigma-T}.}
	\label{perfk24}
\end{figure}

We solve the system of \cref{eq:diff-mass,eq:diff-nu,eq:diff-theta,eq:beta,eq:cutoff} though this time under the more realistic DF given by \cref{eq:FD-DF-cutoff}, with the specific fixed constraints ($\hat N=76.25$, $W_0-\theta_0=24$) and for $mc^2 = \SI{48}{\kilo\eV}$. We solve it for a wide range of control parameters [$\nu_0$, $\beta_0$, $\theta_0$], and plot in \cref{Mk24} all the equilibrium solutions (i.e. extremum of $S$) along the $-\hat M$ vs. $1/\hat T_\infty$ caloric curve as customary.\footnote{
    The plot of a single \textit{astrophysical} caloric curve as \cref{Mk24}, requires very high resolution in order to achieve noise-free spiral features. The boundary condition problem \cref{eq:diff-mass,eq:diff-nu,eq:diff-theta,eq:beta,eq:cutoff} $\&$ \cref{eq:N} is thus solved numerically with a Levenberg-Marquardt least square minimization method, implying a large iterative process (taking $\sim$ few $10^1$ hours of standard desktop CPU-time).
} This problem implies a monoparametric family of solutions, since we have 3 free model parameters (notice that $W_0$ is not independent of $\theta_0$) for two given boundary conditions. We differentiate in \cref{Mk24} among the full family-set of thermodynamically-stable solutions (in continuous-blue line), from the thermodynamically-unstable ones which are shown in dotted-violet. We then analyze in detail all the different kind of density profiles for a fixed value of the total energy $\hat M$ as an example (see vertical dashed line in \cref{Mk24} and \cref{perfk24} for the profiles).

\begin{figure}
	\centering
	\includegraphics[width=\columnwidth]{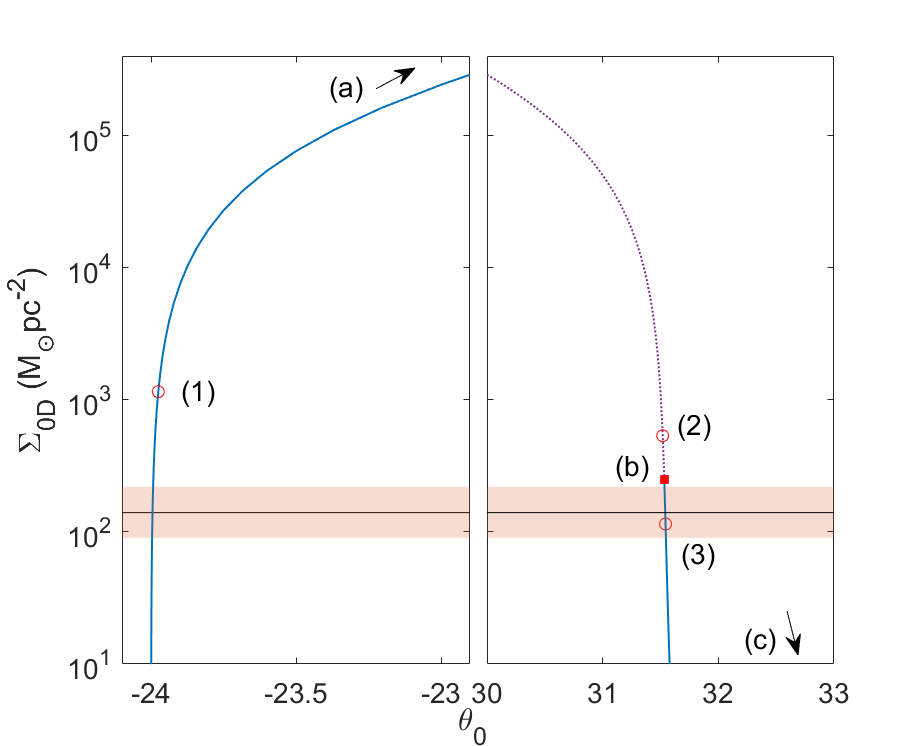}
	\caption{The observationally inferred DM surface density $\Sigma_{0{\rm D}}^{\rm obs} \sim \SI{E2}{\Msun\per\parsec^2}$ (including $3-\sigma$ errors in orange band) \citep{2009MNRAS.397.1169D} is compared with the theoretically prediction $\Sigma_{0{\rm D}}\propto \rho(r_{pl})r_h$, for the full series of equilibrium along the caloric curve of \cref{Mk24}. Only King-like profiles similar to (1) (see \cref{perfk24}) and \textit{core-halo} profiles like (3) are stable and agree with $\Sigma_{0{\rm D}}^{\rm obs}$ at the same time.}
	\label{Sigma0Dk24}
\end{figure}

The main conclusions out of this stability analysis can be summarized as follows:

\begin{enumerate}
    \item Analogously as in \cref{Box2}, the entropy maxima states (either local or global) correspond to solutions with $\theta_0 \ll -1$ lying in the upper continuous-blue branch of \cref{Mk24}, and ending at point (a) where the first instability branch starts. Solutions in this unstable branch (labeled in dotted-violet) are either minimum or saddle points of entropy, and progressively increase their central degeneracy from negative to positive until a \textit{core-halo} profile arises (for $\theta_0 \gtrsim 10$). More precisely, exactly at the point where the spiral start to unwind, a mild degenerate-core with $\theta_0=6.7$ becomes quantum pressure supported being $\lambda_B = 2l_c$. Confirming once more that the Pauli exclusion principle is the responsible of avoiding the monotonic in-spiraling of the caloric curve (i.e. classical gravothermal catastrophe) to occur. Solutions obtained for $10 \lesssim \theta_0 \lesssim 28$, prior to point (b), have a \textit{core-halo} behavior qualitatively similar as the ones obtained in \citealp{2015MNRAS.451..622R}, falling all within the unstable branch. The unstable branch ends at (b), once the caloric curve has rotated as many anticlockwise times as clockwise rotations, and the thermodynamic stability is recovered. Meta-stable solutions in this new stable branch (as solution (3) within the continuous-blue curve) are shown to be of astrophysical interest (see point (iii) below). Such an stability is lost at point (c) when the curve rotates clockwise once again, thus becoming thermodynamically unstable all the way to the second spiral of relativistic origin. The turning-point instability at the left-end of such a spiral, and the concept of gravitational collapse of the core at the last stable point (c), is discussed in detail in \cref{sec:TurningPoint}.
    
    \item The key difference with respect to the \cref{Box2} case is precisely in the second stable branch: it is much more extended in energy and $\hat T_\infty$ and therefore implies a larger family of meta-stable (and stable) states. More importantly, it starts at (b) with central degeneracies $\theta_0\approx 30$ and  $\beta_0\sim 10^{-5}$ typical of astrophysical density profiles \citep{2018PDU....21...82A,2019PDU....24..278A,2020A&A...641A..34B} for such average halo mass. Indeed, the meta-stable solution (3) plotted in \cref{perfk24} is of a perfect astrophysical applicability (see also the item below), since it is similar to that of a Milky-Way RAR DM halo for $mc^2=\SI{48}{\kilo\eV}$, shown to perfectly fit the rotation curve data \citep{2018PDU....21...82A,2019IJMPD..2843003A,2020A&A...641A..34B}. 
    
    \textit{This is a remarkable result, since for the first time a \textit{core-halo} solution like (3) known to be of astrophysical applicability, has now been proven to fall inside the meta-stable branch while being extremely long-lived (as calculated in \cref{sec:appendix:stability_examples}) and thus reachable in Nature. We believe this is not a coincidence: i.e. the fact that in \citet{2018PDU....21...82A,2019IJMPD..2843003A,2020A&A...641A..34B} it was shown the existence of a DM \textit{core-halo} profile where the core can mimic a massive BH while the outer halo can explain the rotation curve, was already a smoking gun for its plausibility in Nature.} 
   % Even more remarkably is the fact that typical central degeneracies $\theta_0\in (30,40)$ developed in this stable branch, are precisely the ones implying central high-density cores which can work as alternatives to the central BH scenario \citep{2018PDU....21...82A,2019PDU....24..278A}.
   
   \item There is a full family of solutions in the second stable branch between (b) and (c) (corresponding to $\theta_0\approx 31.5$ and $\beta_0\in [3.3\times 10^{-5}, 5\times 10^{-5}]$), which are found to be of astrophysical interest: that is, they lie within the allowed DM surface density $\Sigma^{\rm obs}_{0{\rm D}}$ strip as shown in \cref{Sigma0Dk24}, and agree with the expected N-body dispersion velocities ($\sigma_h$) as shown in \cref{Sigma-T}. Interestingly, they cover inner-halo densities at plateau roughly $\rho_p \approx \SIrange{E-3}{E-1}{\Msun\per\parsec^3}$ and total sizes of $R \approx \SIrange{10}{50}{\kilo\parsec}$. In particular, the solution (3) in \cref{perfk24} is the one having the value of $R_{\rm vir} \approx \SI{29}{\kilo\parsec}$ as required from the Press-Schechter analysis within a self-consistent WDM cosmology \cref{sec:appendix:halo-formation}. Finally, solutions with larger $\hat T_\infty$ (i.e. $\beta_0 \gtrsim 5\times 10^{-5}$) towards the relativistic regime approaching point (c) are all of no astrophysical interest: the halos are too diluted and extended to fit within the allowed $\Sigma_{0{\rm D}}$ as explicitly shown in \cref{Sigma0Dk24}.
\end{enumerate}

\paragraph*{Remark} Even if the above conclusions apply for average-sized DM halos, i.e. $M_{\rm vir} \sim \SI{E10}{\Msun}$, with fixed $W_0-\theta_0=24$, we have repeated the stability analysis for other values of $W_0-\theta_0$ between 20 and 28 (for the same $\hat N$).
We have found that for $W_0-\theta_0\lesssim 20$ there are no astrophysical solutions within the meta-stable branch: they acquire inner-halo densities at plateau $\rho_p \gtrsim \SI{1}{\Msun\parsec^{-3}}$ above any reasonable value even for the smallest halos (totally in line with the analysis done in \citet{2019PDU....24..278A}).

Interestingly, for $W_0-\theta_0=24$ there are no stable \textit{core-halo} solutions with a quantum core mass $M_c \lesssim \SI{E7}{\Msun}$. Nevertheless, for larger values of $24 < W_0-\theta_0 < 28$ (and for the same $\hat N$), it is possible to obtain lower (stable) core masses of $\sim \SI{E6}{\Msun}$. The relevance of such (smaller) fermionic core mass is obvious when comparing with our Galaxy (though with a larger $N$ by an order of magnitude than the one here adopted), given it may represent an alternative to the central BH scenario as proven in \citet{2018PDU....21...82A,2020A&A...641A..34B}.
%Indeed, under the choice of $W_0-\theta_0=28$ and for the same $\hat N$, we find the presence of a small family of stable and astrophysical \textit{core-halo} profiles (i.e. belonging to the meta-stable branch somewhat close and below point (b) in the caloric curve) with $M_c \approx \SI{E6}{\Msun}$, confirming (and extending) the findings in \citet{2018PDU....21...82A} for a similar halo mass.

Finally, there exist a threshold value of $W_0-\theta_0$ somewhere between $20$ and $22$, where the meta-stable branch extends only up to an energy value $\hat M$ \textit{smaller} than that of point (a) before becoming unstable (i.e. last stable point (c) is to the left of point (a) in energy). This may imply important consequences regarding the possible gravitational collapse (of gravothermal catastrophic nature) of DM cores towards a BH as can be concluded from the remark in \cref{sec:TurningPoint}. 

\subsection{The $\sigma_h$ - $T$ relation of fermionic halos}

\begin{figure}
	\centering
	\includegraphics[width=\columnwidth]{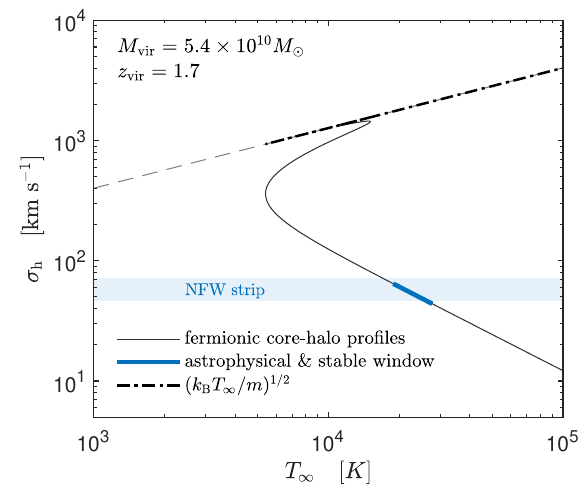}
	\caption{The dispersion velocity $\sigma_h$ of fermionic halos with a \textit{core-halo} morphology lying along \cref{Mk24} is plotted against their effective temperature. It is compared with the traditional isothermal relation $(k_B T/m)^{1/2}$, and with the predicted $\sigma_h$ values arising from N-body simulations within a CDM cosmology. Importantly, it exist a window of $T$ around $10^4$ K where the fermionic \textit{core-halo} profiles are thermodynamically-stable, long-lived, and astrophysically allowed in the sense of both the DM surface density relation $\Sigma_{0{\rm D}}^{\rm obs}$ and the expected $\sigma_h$ from (CDM) N-body simulations.}
	\label{Sigma-T}
\end{figure}

The dispersion velocity of the fermionic halos (calculated as the root-mean-square-velocity at a given halo-scale) is a relevant magnitude which can be compared with the one coming from N-body simulations at virialization. Indeed, within the $\Lambda$CDM cosmology, in \citet{2001ApJ...563..483T} it was calculated a phenomenological expression for the dispersion velocity of the DM halos as a function of the halo mass inside the radius where the rotation curve peaks (dubbed here as $\sigma_h$). Thus in \cref{Sigma-T} we compare the behavior of such a dispersion velocity (labelled in light blue, within the NFW concentration parameters $c$ as reported in \citealp{2001ApJ...563..483T}), with respect to the one of fermionic halos lying along the caloric curve of \cref{Mk24}, and having $M(R_{\rm vir}) \approx \SI{5.4E10}{\Msun}$ occurring at $z_{\rm vir}\sim 2$ (see \ref{sec:appendix:halo-formation} and above section). The fermionic $\sigma_h$ values are plotted as a function of its (effective) temperature $T_\infty \equiv T$, and calculated for tidally-truncated fermionic solutions with a phase-space DF given by \cref{eq:FD-DF-cutoff}. Such a $\sigma_h-T$ relation is explicitly shown in \cref{Sigma-T} for \textit{core-halo} solutions (i.e. $\theta_0\gtrsim 10$) lying along the caloric curve of \cref{Mk24}, the very same relation exist for the branch of diluted (King-like) solutions with $\theta_0 < -1$. The reason for the existence of both branches of solutions with the same $\sigma_h-T$ relation, is because for each \textit{core-halo} solution along \cref{Mk24}, it exist another one in the diluted regime which closely matches the halo part of the former, inside which the dispersion velocity is evaluated (see e.g. the behaviors between solution (1) and (3) in \cref{perfk24}).
%The fermionic dispersion velocity plotted in fig. is evaluated at the maximum of the halo rotation curve $r_{max}$, in order to make a one-to-one comparison with the prediction obtained from N-body simulations at virialization as originally reported within a CDM cosmology in (Taylor-Navarro). 

One important result from this analysis is the fact that the $\sigma_h-T$ of fermionic halos \textit{does not} follow the traditional isothermal trend $\sigma \propto (k_B T/m)^{1/2}$. Instead, it deviates from it according to the different behavior of the caloric curve of \cref{Mk24} in which the temperature covers a wide range of regimes, from relatively hot to relatively cold, depending of the central degeneracy and cut-off parameter (the latter two not present in the traditional isothermal scenario). That is, for \textit{core-halo} solutions starting with $\theta_0=10$ in the caloric curve (just after the spiraling out at the top of \cref{Mk24}), the temperature increases with corresponding increase of $\sigma_h$ closely following the Boltzmannian relation. This trend continues until the first anti-clockwise turn of the caloric curve (in the inner part of the location of point (a)), where $T$ decreases (and so does $\sigma_h$). This second trend ends when the caloric curve reaches the maximum (from the inside curve), and so $T$ starts to rise once more (in this case with decreasing $\sigma_h$), all the way until the onset of instability at point (c).

The main conclusion from this analysis is that: \textit{it exist a window of effective $T\sim$ few $\SI{E4}{K}$ falling within the range of thermodynamically stable core-halo solutions, with corresponding $\sigma_h$ values which roughly agree with the predicted window arising from N-body simulations within a CDM cosmology as given in \citet{2001ApJ...563..483T}.} Interestingly, such very same range of $T$, practically coincides with the one of the astrophysical family of solutions (shown in thick-blue in \cref{Sigma-T}) in the sense of the DM surface density relation of \cref{Sigma0Dk24}.

This dispersion velocity analysis allow us to \textit{link} the temperature of the DM fermions prior and after virialization, and to know which values are the realistic ones (for given particle mass) in the sense of the expected $\sigma_h$ from simulations. Indeed, while typical DM redshifting temperature of the fermions just prior halo formation is about few Kelvin (i.e. at $z_{\rm vir}=10$ as calculated from \ref{sec:appendix:halo-formation} for resonantly produced sterile neutrinos), the effective $T$ of the very same particles in the fermionic halos of $\sim \SI{5E10}{\Msun}$, must be $\sim \SI{E4}{K}$. That temperature gap between prior and after relaxation, can be understood in terms of the violent relaxation mechanism. It mixes the fermion gas and makes the (effective) temperature of the quasi relaxed halo hotter as expected from the negative specific heat acting on this kind of self-gravitating systems.
%The reason applied for CDM, though the $\sigma$ gap in our approach should be lees abrupt than in the CDM case, since in the latter the $\sigma$ prior to formation can be considered as $0$.

\subsection{Comparison with other works}
\label{sec:comparison}
We start by emphasizing that the stability results presented in this section are completely original, since it is the first time they are obtained for tidally-truncated self-gravitating fermions at finite temperature in GR, with realistic boundary conditions at virialization. Nevertheless, it is important to compare the main conclusions obtained in \cref{sec:cutoff-case}, with those of a similar stability analysis done for self-gravitating fermions within Newtonian gravity in \citet{2015PhRvD..92l3527C} (see \cref{sec:appendix:stability_examples} for comparisons with other works using the box-confined ansatz for fermions in GR).

In fig. 30 of that work, Chavanis et al. obtained a similar caloric curve as the one obtained here in \cref{Mk24}, though without the second spiral of relativistic origin given he worked in Newtonian gravity. The importance of this comparison, is that both caloric curves have the similar features: both  (microcanonical) stables branches are of similar features and are both applicable to relatively large halos. Indeed, in \citet{2015PhRvD..92l3527C} they did the analysis for a dimensionless parameter $\mu=\num{E9}$ (\textit{not} the chemical potential), which it can easily shown to be equal to ${\rm e}^{W_0-\theta_0}$, thus implying a value for $W_0-\theta_0\approx 21$ close to the one chosen in \cref{sec:cutoff-case}. 

In fig. 45 of that work they provided all the possible density profiles at a given energy, similarly as done here in \cref{perfk24}, though they did for an energy value already within the spiral feature. They showed basically three different kind of profiles: (i) a King-like profile belonging to the first stable branch dubbed as the \enquote{gaseous-phase} (similar to solution (1) in \cref{perfk24}); (ii) a \textit{core-halo} profile belonging to the unstable branch dubbed as the \enquote{embrionic-phase}; and (iii) another kind of \textit{core-halo} profile belonging to the second stable branch and dubbed by Chavanis et al. as the \enquote{condensed-phase} (similar to solution (3) in \cref{perfk24}). 

Up to this point both qualitative results about the thermodynamic stability (i.e. the one given in \citealp{2015PhRvD..92l3527C} and the one from \cref{sec:cutoff-case}) are somewhat in line, though Chavanis et al. attempted a very different conclusion with respect to the one obtained here regarding the applicability to DM halos. They concluded that (I) either you have \textit{core-halo} solutions belonging to the \enquote{embrionic-phase} (qualitatively similar as the ones obtained in \citealp{2015MNRAS.451..622R}) but are thermodynamically unstable (i.e. unreachable), or (II) you get stable \textit{core-halo} profiles belonging to the \enquote{condensed-phase}, but cannot explain the DM content in large galaxies since the halo is too extended and diluted for such a goal. While we generally agree on conclusion (I) as shown in \cref{sec:box-case} and \cref{sec:cutoff-case}; we totally disagree with conclusion (II). Moreover, we have proven in \cref{sec:cutoff-case} that such a conclusion is indeed wrong. 
% \footnote{
    %Indeed, we have reached the same conclusion here (and in \cref{sec:box-case}), given the so-called \enquote{embrionic-phase} profiles in Chavanis et al. are obtained here as well for $10 \lesssim \theta_0 \lesssim 28$, falling all within the unstable branch.
%}

Such a discrepancy in the interpretation of the results is understood when introducing a proper quantitative and dimensionfull analysis of the profiles in relation to DM halo observables, together with the quest for a full coverage of the energy values $\hat M$ along the metastable branch (not developed in \citealp{2015PhRvD..92l3527C}). That is, for a particle mass $mc^2 \sim \SI{50}{\kilo\eV}$ there exist an accessible energy window $\hat M$ in which there are meta-stable (and long-lived) \textit{core-halo} profiles which acquire the observationally inferred inner halo densities ($\sim \SI{E-2}{\Msun\per\parsec^3}$) and virial radii ($\sim$ few $\SI{10}{\kilo\parsec}$) typical of average size galaxies. Indeed, this observational correspondence is evidenced through the $\Sigma_{0{\rm D}}$ relation shown in \cref{Sigma0Dk24} and further explained in \cref{sec:cutoff-case}. 

This kind of thermodynamically stable \textit{core-halo} profiles correspond with the so-called \enquote{condensed-phase} introduced in \citet{2015PhRvD..92l3527C}, and has been already implemented in \citet{2018PDU....21...82A,2019IJMPD..2843003A} to provide an excellent fit to the Milky Way rotation curve. Remarkably enough, the degenerate core of this last kind of solutions (for particle masses in the range  $\sim \SIrange{E1}{E2}{\kilo\eV}$) can mimic the massive BH in SgrA* \citep{2018PDU....21...82A,2020A&A...641A..34B} as well, a result which \textit{is not} possible for \enquote{embrionic-phase} solutions as shown in \citet{2015MNRAS.451..622R} (by the way unstable). 

\paragraph*{Remark} In \citet{2015PhRvD..92l3527C}, the highly degenerate core (i.e. $T \to 0$) belonging to the \enquote{condensed-phase} kind of stable solutions, were used as a potential candidate to explain the DM halos in dwarf galaxies for particle masses $\sim \SI{1}{\kilo\eV}$ (as motivated from the results in \citealp{2013NewA...22...39D}).\footnote{
    More refined phenomenological analysis under such fully degenerate regime, indicate that sub-\si{\kilo\eV} fermion masses are needed to provide decent fits to dispersion velocity data in dSphs \citep{2015JCAP...01..002D}.
} However, we notice that from the discussion above, there is a priori no necessity to go to such low fermion masses of $\sim \SI{1}{\kilo\eV}$ (or below) in order to explain the DM halos in dwarf galaxies. In other words, it is absolutely possible to provide good fits to dispersion velocity data in such galaxy types, using the overall stable \textit{core-halo} profiles (like solution (3) in \cref{perfk24} but for lower $M_{\rm vir} \sim \SI{E8}{\Msun}$), as shown in \citet{2019PDU....24..278A}. We thus claim, in views of the results here presented together with the phenomenology for DM halos made in \citet{2018PDU....21...82A,2019PDU....24..278A,2020A&A...641A..34B}, that the semi-degenerate fermion regime (i.e. leading either to diluted or \textit{core-halo} profiles) is enough to explain the plethora of DM halos without the need to invoke the (extreme) fully degenerate ($T \to 0$) regime.

Moreover, such fully-degeneracy paradigm for $mc^2\lesssim \SI{1}{\kilo\eV}$ aimed to be applicable to dSph DM-halos, suffers from many problems or tensions such as: (a) Ly $\alpha$ forest constraints \citep{2017JCAP...06..047Y}, (b) phase-space bounds and MW satellite counts \citep{2014PhRvD..89b5017H}, and even (c) dispersion velocity fits in dwarfs since an \textit{extra} isothermal halo component has been shown in \citet{2017MNRAS.467.1515R} to be needed in order to agree with data. Problems which naturally disappear within our $\mathcal{O}(\SI{10}{\kilo\eV})$ WDM paradigm. 

\section{Gravitational collapse and turning point instability}
\label{sec:TurningPoint}

Fermionic \textit{core-halo} solutions with $N > N_{\rm OV}$ at virialization, may eventually become unstable (either thermodynamically and dynamically) and undergo a gravitational core-collapse as we show below. Historically, the gravitational collapse of a degenerate and relativistic \enquote{star} at a specific central density $\rho_0$, was understood in terms of the onset of thermodynamical (and dynamical) instability at a turning-point \citep{1965gtgc.book.....H,1981ApJ...249..254S,1988ApJ...325..722F}. Such a turning-point (TP) being defined as the point where the total mass is a maximum respect to $\rho_0$, i.e. $\rm dM/\rm d\rho_0=0$. Importantly, in  \citet{1981ApJ...249..254S,2014CQGra..31c5024S} it was demonstrated that for any EoS (e.g. not necessarily isentropic) the existence of a TP along a smooth sequence of GR equilibrium states, implies the presence of a thermodynamic instability on one side of the TP.     

However, turning-points do not provide a necessary condition for thermodynamic instability, and the onset of such an instability could occur even without the existence of a TP at all (see point (a) in \cref{fig:calcurvebox1} with $N<N_{\rm OV}$ for an explicit example, and \citealp{2014CQGra..31c5024S} for a general theoretical result). Moreover, the onset of thermodynamical (and dynamical) instability can occur \textit{prior} to the TP in the $\rho_0$ vs. $M$ curve, as first shown numerically in \citet{2011MNRAS.416L...1T} for rotating perfect-fluid \enquote{stars} (contrary to what historically expected in \citealp{1988ApJ...325..722F}). 

\begin{figure}
	\centering
	\includegraphics[width=\columnwidth]{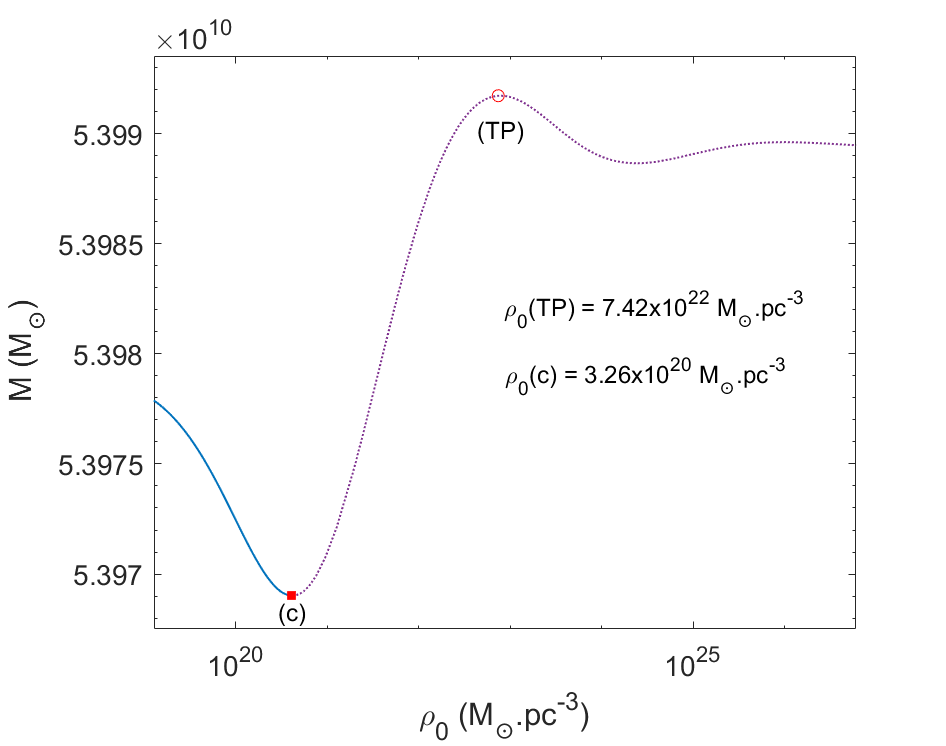}
	\caption{Series of equilibrium states with $N > N_{\rm OV}$ are shown along a $\rho_0$ vs. $M$ curve, in correspondence with the relativistic and degenerate end of the caloric curve of \cref{Mk24}. At difference with standard degenerate fermion 'stars', the last stable configuration (c) of self-gravitating fermions at finite $T_\infty$, occurs just at the minimum of the curve and \textit{prior} to the TP instability.}
	\label{MtotVsrho0}
\end{figure}

Importantly, for the first time we confirm such a conclusion but for a perfect-fluid (neutral) fermionic non-rotating \enquote{star}, with an EoS including for temperature effects as demonstrated below and in \cref{MtotVsrho0,Mk24}.   

\begin{figure}
	\centering
	\includegraphics[width=\columnwidth]{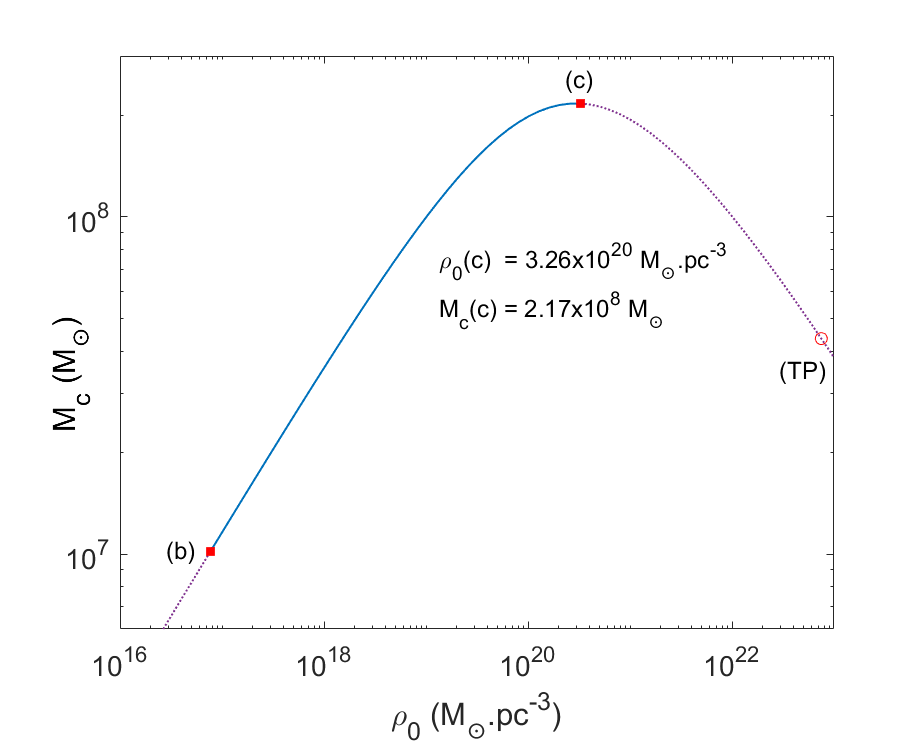}
	\caption{The last stable configuration (c) at the onset of the core-collapse (corresponding with point (c) in \cref{Mk24}), is acquired at the maximum of a central density vs. \textit{core} mass, and having a critical core mass $M_c^{\rm cr}\approx \SI{2E8}{\Msun}$ for $mc^2=\SI{48}{\kilo\eV}$. The core mass at the TP is nearly an order of magnitude below the critical mass.} 
	\label{McVsrho0}
\end{figure}

Moreover, we are able to localize the TP instability (i.e. the maximum in a $\rho_0$ vs. $M$ curve as in \cref{MtotVsrho0}), in the left-lower end of the caloric curve, where the spiral of relativistic origin rotates clockwise (see the empty red-circle in \cref{Mk24}). Clearly, such TP occurs at a different energy respect to the last stable configuration (c) along the unstable branch of the caloric curve. 

Thus we have proven, that for self-gravitating systems at finite $T_{\infty}$ in GR, the TP instability \textit{does not} coincide with the last stable configuration occurring at (c). This original result bares an important consequence regarding the concept of critical mass for gravitational core-collapse $M_c^{\rm cr}\propto m_{p}^3/m^2$ where $m_{p}$ is the Planck mass,\footnote{
    See \citet{2014IJMPD..2342020A} for a numerical demonstration in GR on the finite $T$ effects in the critical core mass $M_c^{\rm cr}$.
} traditionally explained in terms of the TP instability for fully degenerate stars (see e.g. \citealp{1983bhwd.book.....S}). Such a correspondence \textit{does not} apply here, and instead, the $M_c^{\rm cr}$ value is achieved at the last stable configuration (c) placed at the \textit{minimum} of the $\rho_0$ vs. $M$ curve, thus occurring \textit{prior} to the TP as can be directly checked by comparing \cref{Mk24,MtotVsrho0}.

Interestingly, such a last stable configuration acquiring the critical core mass, occurs instead at the maximum, but in a $\rho_0$ vs. $M_c$ curve as shown in \cref{McVsrho0}. Indeed, the value of the core mass at the TP is near an order of magnitude below the critical $M_c^{\rm cr}$ as can be seen directly by comparing \cref{McVsrho0,MtotVsrho0}. The right value of the critical mass $M_c^{\rm cr}$ (i.e. the one associated with the last stable configuration and not with the TP), is of central importance for super massive BH formation and astrophysics in general. As a clear example, for $mc^2 = \SI{48}{\kilo\eV}$ it is possible to form a SMBH of $M_c^{\rm cr}\approx \SI{2E8}{\Msun}$ (see \cref{McVsrho0}) at the center of a realistic DM halo as explained in \cref{sec:cutoff-case} and in the remark below. The relevance of such a DM core-collapse scenario, is that it can occur in the high redshift ($z\sim 10$) Universe at halo virialization (see \cref{sec:appendix:halo-formation}), without the need of prior star formation, or other BH seed mechanisms involving super-Eddington accretion rates. That is, the thermodynamics of tidally-truncated self-gravitating fermions in a cosmological setup, can offer a powerful tool for SMBH formation worthy for further investigation.  

We have further calculated the central redshift ($z_0=e^{-\nu(0)/2}-1$) of a light source at rest at the center of a DM core (for all the solutions along the caloric curve of \cref{Mk24}), as a measure of how relativistic they are. For the critical solution (c) it gives $z_0^{cr}=0.478$ (at $\hat T_{\infty}=5.11\times 10^{-3}$), while for the TP (unstable) core it gives $z_0^{TP}=3.16$ (at $\hat T_{\infty}=4.24\times 10^{-3}$). The arising of thermodynamically (and dynamically) unstable (i.e. collapsing to a BH) solutions for $z_0^{cr}\gtrsim 0.5$ found here for tidally truncated configurations of keV fermions (with $N>N_{OV}$ in the high $T$ regime), is in line with former results found in \cite{1989ApJ...344..146R} for relativistic clusters.

\paragraph*{Remark 1} Since we do not solve the time-evolution of the fermionic configurations once they become thermodynamically (and dynamically) unstable, the concept of gravitational collapse deserves further explanation. In the caloric curves with $N>N_{\rm OV}$ under study, there are two kind of possible gravitational collapses which may arise. For a system starting at virialization in the diluted (stable) regime before point (a), one can think that such a state \textit{evolves} quasi-stationary in time while looses energy due to evaporation and (long-range) collisions, until reaching the threshold energy of point (a). At this critical energy, the configuration will evolve directly towards the next accessible stable-state just below (a) in the meta-stable branch, and leading to the new \textit{core-halo} configuration. Such limiting phase-transition, i.e. from diluted to a semi-degenerate one occurring at such critical energy, is usually called as a collapse of gravothermal catastrophic nature \citep{2019arXiv190810303C}. 

If it keeps loosing energy along the second stable branch, it will eventually reach the last stable configuration at (c) (see \cref{Mk24}), below which there is no possible accessible state, and the system must collapse towards a massive BH (according to the core-collapse criterion of relativistic origin explained above). For a further discussion including the different time-scales involved between the two different collapsing processes: i.e. between the gravothermal catastrophe at point (a), and the gravitational collapse of relativistic origin occurring at (c), see \citet{2019arXiv190810303C} and references therein. 
% \footnote{
    %This is at difference with the analogous energy-loosing process for $N<N_{\rm OV}$, where there always exist an equilibrium state for all the accessible values of energies and $T_\infty$ \citep{2015PhRvD..91f3531C}
%}

\paragraph*{Remark 2} We notice that a complementary mathematical proof about the onset of thermodynamical (and dynamical) stability in the microcanonical ensemble, may be obtained by explicitly calculating the second order variations of entropy and their sign changes, following the work of \citet{2013CQGra..30k5018R}. Though, even if in that work it was explicitly calculated the expressions for $\delta^2 S$ for a self-gravitating system of particles in GR under a perfect fluid assumption, it was done only for barotropic equations of state. The fermionic equations of states used in this paper are of more general form, written only parametrically, thus more sophisticated calculations are needed in the present case (if at all possible) in order to attempt such a complementary proof.

%\clearpage

\section{Conclusions}
\label{sec:conclusion}

We have studied the formation and stability of collisionless self-gravitating system of fermions at finite $T$ in GR, with applications to the problem of virialization of DM halos in a realistic cosmologial framework. Unlike the N-body numerical simulation-approach, we have assessed these issues by means of a thermodynamical approach for self-gravitating fermions, which eventually maximize its coarse-grained entropy. In particular we have performed a thermodynamic stability analysis in the microcanonical ensemble for solutions with given particle number $N$, at the moment of halo-virialization in a WDM cosmology, and further calculate the life-time of the metastable equilibrium states. 
% \textbf{The full methodology here adopted is: i) we calculate the linear power spectrum for the fermionic candidates; ii) we use the corresponding Press-Schechter formalism to obtain the virial halo mass, $M_{\rm vir}$, with associated redshift $z_{\rm vir}$; iii) by assuming a maximum entropy production principle for halo formation,  we obtain the full family (from King-like to  \textit{core-halo}-like) of such fermionic DM profiles in agreement with above virial constraints; iv) we calculate the stability and typical life-time of such equilibrium states within a thermodynamical approach for solutions with given particle number in the microcanonical ensemble.}

The advantages of our numerical approach is that it allows for a detailed description of the relaxed halos from the very center to periphery, not possible in N-body simulations due to finite inner-halo resolution. In addition, it includes richer physical ingredients such as (i) General Relativity --- necessary for a proper gravitational DM core-collapse to a SMBH; (ii) the quantum nature of the particles --- allowing for an explicit fermion mass dependence in the profiles; (iii) the Pauli principle self-consistently included in the phase-space DF --- giving place to novel \textit{core-halo} profiles at (violent) relaxation. 

Our approach allows to link the behavior and evolution of the dark matter particles from the early Universe all the way to the late stages of non-linear structure formation at virialization. That is, we start by calculating the linear matter power spectrum for a $\mathcal{O}(\SI{10}{\kilo\eV})$ DM sterile neutrino, to then use the corresponding Press-Schechter formalism to obtain the virial halo mass, $M_{\rm vir}$, with associated redshift $z_{\rm vir}$ (see \cref{sec:appendix:halo-formation}). The fermionic halos are assumed to be formed by fulfilling a maximum entropy production principle at virialization. It allows to obtain a most likely DF of Fermi-Dirac type as first shown in \citet{1998MNRAS.300..981C} (generalizing Lynden-Bell results), which is here applied to explain DM halos. Moreover, such a DF is used here to calculate the full family (from King-like to  \textit{core-halo}-like) of fermionic equilibrium profiles in full GR, in agreement with prior virial constraints. Finally, the stability, typical life-time of such equilibrium states, as well as their possible astrophysical applications are studied within a thermodynamic approach.  

The disadvantages of our procedure, is that it does not explicitly include for the accretion and merger processes in a time dependent manner as achieved in numerical simulations. However, we recall that the violent relaxation mechanism (as the one applying here within a maximum entropy production assumption), takes place subsequent times at each merging process event, probabilistically included in the Press-Schechter formalism through the mass variance $\sigma(M)$.  

%\textbf{However, TALK ABOUT THE MERGING HISTORY EFFECTS CAUSED BY MASSIVE SATELLITES (AS LMC) ON THE STABILITY OF OUR MILKY WAY HALO (from recent results Erkal et al. 20201). AND HOW MINOR SUCH EFFECTS IMPACTS ON OUR FREE PARAMETERS AND MASS ESTIMATIONS.}

Our approach is self-consistent, in the sense that the nature and mass of the DM particle involved in the linear matter power spectrum calculations (obtained within a CLASS code for WDM), is the very same building block at the basis of virialized DM configurations with its inherent effects in the \textit{core-halo} profiles. It applies to spherical and rather isolated DM configurations which just underwent a violent relaxation process within a $\mathcal{O}(\SI{10}{\kilo\eV})$ WDM cosmology. Such configurations can start forming in the high $z\sim 10$ Universe, though take place more ubiquitously at $z\sim 2$, with boundary halo conditions consistent with a Press-Schechter theory of non-linear structure formation (see \cref{sec:appendix:halo-formation}).

We outline below all the main theoretical results and its astrophysical consequences obtained in this work. 
%In \ref{conclusion:box} we summarize the conclusions when assuming a self-gravitating system of fermions distributed in phase-space by a Fermi-Dirac DF (\cref{eq:FD-DF}), which is contained in a spherical-box in order not to cause an (coarse-grained) entropy runaway. Then in \ref{conlusion:cutoff} we summarize the main result of this work dealing with the (more realistic) tidally-truncated system of fermions distributed in phase-space by \cref{eq:FD-DF-cutoff}. In \ref{conclusion:consequences} we briefly point out the astrophysical and cosmological consequences of our results regarding SMBH formation, and in \ref{conclusion:profiles} and \ref{conclusion:critical-mass} we describe the new theoretical results dealing with the TP instability and the shape of the fermionic caloric curves in GR. 

\begin{enumerate}
    \item \label{conclusion:box} \textit{Among all the GR spherically-symmetric self-gravitating systems of $\mathcal{O}(\SI{10}{\kilo\eV})$-fermions confined in a spherical-box, which maximize the coarse-grained entropy at halo-virialization in a WDM cosmology, do not exist any thermodynamically-stable \textit{core-halo} configuration with halo masses $\sim \SIrange{E9}{E10}{\Msun}$, able to agree with the observed DM surface density relation $\Sigma_{0{\rm D}}$. Instead, diluted-Fermi configurations (resembling pseudo-isothermal spheres) do fulfill with both the thermodynamic stability and the DM halo $\Sigma_{0{\rm D}}$ phenomenology in the above DM halo mass range.}\footnote{
        This statement is expected to hold for any halo mass above $\sim \SI{E9}{\Msun}$, since the central degeneracy parameter where meta-stability sets in (after point (b)) is larger for larger total masses, thus implying \textit{core-halo} profiles with even more extended and diluted halos clearly disfavoured by data.
    }. See \cref{sec:box-case} for details. 
    \item \label{conlusion:cutoff} \textit{Among all the GR spherically-symmetric and tidally-truncated self-gravitating systems of $\mathcal{O}(\SI{10}{\kilo\eV})$-fermions, which maximize the coarse-grained entropy at halo-virialization in a WDM cosmology, it exist thermodynamically-metastable \textit{core-halo} configurations which are long-lived and agree with the observed DM surface density relation $\Sigma_{0{\rm D}}$, and with the expected N-body dispersion velocities ($\sigma_h$) for $\sim \SI{E10}{\Msun}$ halos}. Such kind of stable \textit{core-halo} fermionic profiles have effective temperatures of few $\SI{E4}{K}$, and are precisely of the same kind as the ones applied recently in \citet{2018PDU....21...82A,2019PDU....24..278A,2020A&A...641A..34B} to explain the rotation curve data in galaxies, with the DM core able to mimic the super massive BHs at their centers. See \cref{sec:cutoff-case} for details. 
    \item \label{conclusion:consequences} The thermodynamic formalism for self-gravitating $\mathcal{O}(\SI{10}{\kilo\eV})$-fermions introduced here, allows for a SMBH formation mechanism through the DM core-collapse of relativistic origin (see \cref{sec:TurningPoint}). Interestingly, it can start within the high $z\sim 10$ Universe, without the need of prior star formation or any BH seed mechanisms involving (likely unrealistic) super-Eddington accretion rates. More generally, a dense quantum core (i.e without the singularity) at the center of a stable and average-sized DM halo, can reach masses between $\sim \SIrange{E6}{E8}{\Msun}$ (see \cref{sec:cutoff-case}) which may provide an alternative for the traditional SMBH scenario \citep{2018PDU....21...82A,2019PDU....24..278A,2019IJMPD..2843003A,2020A&A...641A..34B}. 
    \item \label{conclusion:profiles} We have calculated for the first time the caloric curves for self-gravitating tidally-truncated $\mathcal{O}(\SI{10}{\kilo\eV})$ fermions at finite $T$ within GR, and applied to realistic DM halos (i.e. sizes and masses). Our results confirm and extend the double-spiral feature (the first of quantum nature and the second of relativistic origin) in the caloric curves with fixed $N$ as recently obtained in \citet{2020EPJB...93..208A}. With the precise shape of the caloric curves, we have applied the Katz criterion for thermodynamic stability (\cref{sec:appendix:stability_examples}), finding different families of stable as well as astrophysical DM profiles. They are either King-like (similar to Burkert), or develop a \textit{core-halo} morphology able to fit the rotation curve in galaxies \citep{2018PDU....21...82A,2019PDU....24..278A}. In the first case the fermions are in the dilute regime (i.e. $\theta_0\ll-1$) and correspond to a global maxima of entropy, while in the second case, the degeneracy pressure (i.e. Pauli principle) is holding the quantum core against gravity, and correspond to a local maxima of entropy. Such meta-stable states are extremely long-lived as shown in \cref{sec:appendix:stability_examples}, and more likely to arise in Nature than the former as argued in \citealp{2005A&A...432..117C}).      
    \item \label{conclusion:critical-mass} We proved for the first time that for tidally-truncated self-gravitating systems of neutral fermions at finite $T$ in GR, the thermodynamical (and dynamical) instability occur \textit{prior} to the TP in the $\rho_0$ vs. $M$ curve, as explicited in \cref{sec:TurningPoint}. Indeed, the critical mass of gravitational core-collapse $M_c^{\rm cr}$ is achieved at the last stable configuration (with lower energy with respect to the TP), which interestingly coincides with the maximum but in a \textit{core mass} $M_c$ vs. $\rho_0$ curve. Given the value of $M_c$ at the TP can differ by an order of magnitude below the real $M_c^{\rm cr}$, it shows the importance of this result regarding the SMBH mass estimates, when applied to astrophysics. 
\end{enumerate}

The DM fermion mass of $\mathcal{O}(\SI{10}{\kilo\eV})$ used in this work produce, down to Mpc scales, the same $\Lambda$CDM power-spectrum, hence providing the expected large-scale structure \citep{2009ARNPS..59..191B}. Moreover, since the fermion mass is larger than $>\SI{5}{\kilo\eV}$, it is not in tension with constraints from the Lyman-$\alpha$ forest \citep{2009PhRvL.102t1304B,2013PhRvD..88d3502V,PhysRevD.96.023522}, nor with the number of Milky Way satellites \citep{2008ApJ...688..277T}.  

To conclude, we believe the results shown in this paper may open new insights in the formation and evolution of galaxies. Moreover, the degeneracy-pressure-supported core at the center of the stable DM profile, and its eventual core-collapse, may play crucial roles in helping to understand the formation of SMBHs in the high $z$ Universe, or in mimic its effects without the need of the singularity at all. The astrophysical consequences of the analysis here developed --- together with the results recently presented in \citealp{2018PDU....21...82A,2019PDU....24..278A,2020A&A...641A..34B}) --- strongly suggest that such DM \textit{core-halo} morphologies may be a plausible scenario within the late stages of non-linear structure formation, which should start to be seriously considered in the field.

\section*{ACKNOWLEDGMENTS}
The authors thank G. V. Vereshchagin for a critical reading of this paper, and to J. A. Rueda, C. Llinares and C. A. Vega-Mart\'inez for useful discussions about different aspects of the work. C.R.A has been supported by CONICET and MINCYT (code: PICT-2018-03743), and Secretary of Science and Technology of FCAG and UNLP. M.I.D has been partially supported by the University of Buenos Aires, and by the ANR project MOMA (France). R.Y was supported by La Sapienza University of Rome and ICRANet.

\section*{DATA AVAILABILITY}
The data underlying this article will be shared on reasonable request to the corresponding author.

\bibliographystyle{mnras}
\bibliography{SecPS}

\appendix
\section{Thermodynamic stability criterion $\&$ lifetimes of metastable states}
\label{sec:appendix:stability_examples}

\subsection{Thermodynamic stability: the Katz criterion}
\label{sec:Katz}

The thermodynamic stability analysis pertinent to this work (valid either in the microcanonical or the canonical ensembles) is carried out following the criterion described by Katz in \citet{katz1978number,katz1979number}. This is a powerful method based on the theory developed by Poincaré \citep{1885Pincare}, which allows to obtain the number of unstable modes only from the topological properties of the series of equilibrium, without the need to calculate the full eigenvalue problem of the perturbed system. In this section we first summarize the method in a rather generic and formal manner, and then we provide a \enquote{rule of thumb} on how to apply it in an easy way depending on the ensemble under consideration.

Following \citet{katz1978number,katz1979number}, let $f$ be some relevant function of the configuration that finds itself at an extrema, say a maximum, when the system is in a stable equilibrium, and $x$ as a parameter that runs continuously through the series of equilibria. Then, it is there demonstrated that changes of stability for any individual perturbative mode will occur, at a given point in the series, if and only if two specific conditions are met:
\begin{enumerate}
    \item the slope of the $\partial_x f$ vs $x$ curve is infinite (i.e. the tangent is a vertical line), and
    \item the sign of said slope shifts at that point.
\end{enumerate}
Note that this will usually mean the presence of a multivalued $\partial_x f$.

A vertical tangent at a specific point in the series is equivalent to a mode eigenvalue being zero \citep{katz1978number,katz1979number}. Near that point, the sign of the eigenvalue is equal to that of the slope. A positive (negative) eigenvalue means that the system is stable (unstable) regarding perturbations in that mode. It is then immediate that when conditions (i) and (ii) are met, a shift in stability occurs at the single mode level.

A configuration is considered to be at a stable equilibrium whenever all of the modes are stable. Conversely, instability of a single mode suffices to make the whole configuration unstable. Therefore, knowledge of the total number of unstable modes for one single equilibrium is sufficient to determine the stability of all the other equilibria in the family. The procedure thus consists in simply locating a known stable point in the $\partial_x f$ vs. $x$ diagram, and then following the curve, locating the points where conditions (i) and (ii) are met, while counting the resulting number of unstable modes.

When working in the microcanonical ensemble, it is natural to define $f$ as the entropy of the system ($f\equiv S$) and $x$ as the thermodynamic parameter of the ensemble, the energy, which in GR is equal to the (dimensionless) total mass of the system ($x\equiv \hat M$). This yields the derivative $\partial_x f \equiv \frac{\partial S}{\partial \hat M}=\hat T_{\infty}^{-1}$ (the inverse temperature). Thus, the relevant curves in this context must be displayed through a $T_{\infty}^{-1}$ vs. $M$ plot. Though, in order to keep with conventions and more easily compare with other works, we choose to plot $\hat T_{\infty}^{-1}$ vs. $-\hat M$ instead throughout the paper. This simple reverses the meaning of the sign of the slope near a vertical turning point; a negative (positive) slope now means a stable (unstable) mode.

Following the explained above (as e.g. in the microcanonical ensemble) as well as \citet{1998MNRAS.296..569C,2015PhRvD..92l3527C}, we can state as a practical \enquote{rule of thumb} that: (a) the arising of an unstable mode (when the negative slope in the $T_{\infty}^{-1}$ vs. $-M$ curve becomes infinite for then turning into positive) is equivalent to say that the caloric curve \enquote{rotates clockwise} and viceversa, (b) when the same curve \enquote{rotates anticlockwise} implies that a stable mode has been re-gained (the later implying that a positive slope turned into negative just after becoming vertical). In this sense, once in a given unstable branch of the caloric curve (coming from an originally stable branch), it is necessary as many anticlockwise turns of the curve as clockwise passed, to regain the thermodynamic stability. 

\begin{figure}
     \centering
     \includegraphics[width=\columnwidth]{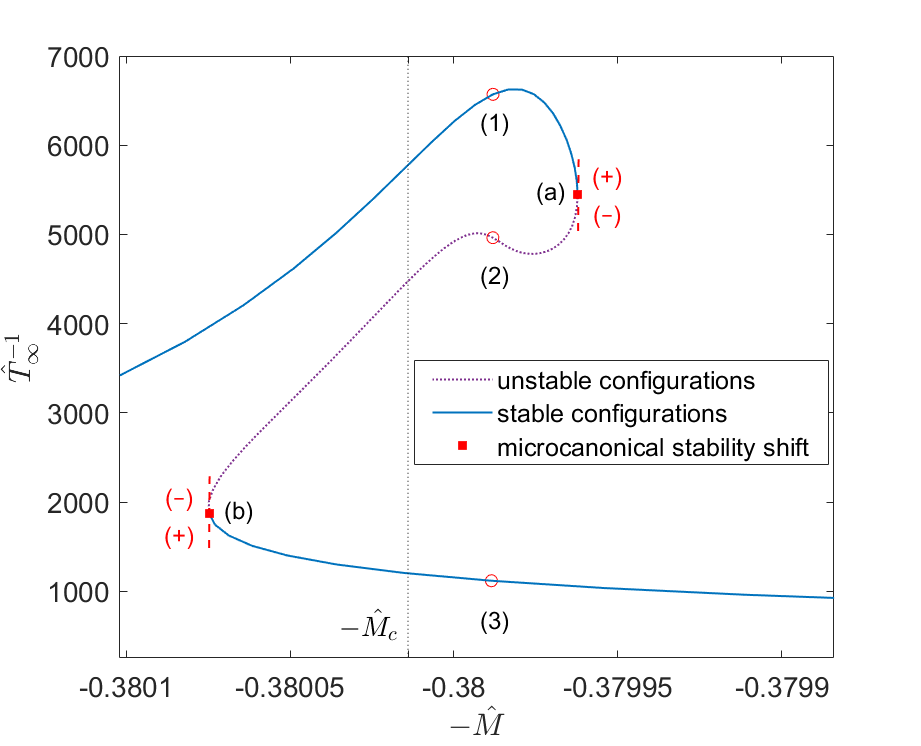}
     \caption{Series of equilibrium solutions along the caloric curve for box-confined configurations of fermions with $\hat N=0.38 < N_{\rm OV}$ and $\hat R=10^3$. For such relatively small value of $\hat R$ the curve develops a \enquote{dinosaur's neck}, where the gravothermal catastrophe typical of Boltzmannian systems is avoided thanks to central degeneracy, as first shown in Newtonian gravity in \citet{1998MNRAS.296..569C}. Every time the curve rotates clockwise (point (a) in the microcanonical ensemble), it losses a stability mode, which is re-gained when rotates anticlockwise (at point (b)) according to the Katz criterion of thermodynamic stability. The stable (unstable) branches in relation with the entropy maxima (minimum/saddle points) is shown in \cref{S_a10}.}  
     %The states within the continuous-blue branches are thermodynamically (and dynamically) stable (i.e. either local or global entropy maxima), while the dotted-violet branch - between (a) and (b)- is unstable (i.e. either minimum or saddle point of entropy), according to the stability criterion of \cref{sec:appendix:stability_examples}.}
    \label{ej_M_c3}
\end{figure}

We exemplify this process for a box-confined case (i.e. $\hat N=0.38 < N_{\rm OV}$ and $\hat R=1000$) in \cref{ej_M_c3}, where the stable branches of solutions are plotted in continuous-blue, and the unstable branches are displayed in dotted-violet. A qualitative analysis in the microcanonical ensemble which is sequential in its nature, proceeds as follows. When $\theta_0 \ll -1$, the systems behaves like a classical Boltzmannian self-gravitating gas, robust in its stability \citep{1968MNRAS.138..495L}. Such solutions lay in upper continuous-blue curve identified as the stable branch, up until the first clockwise turn takes place (i.e. gaining an instability mode see \cref{ej_M_c3}). At this point the caloric curve spirals inwards in the unstable branch, corresponding with the semi-degenerate regime of solutions when $\theta_0$ is just turning into positive. For this small $\hat R$ case, this trend continues until the curve rotates anti-clockwise and the stability mode is re-gained.\footnote{
    For much larger system sizes (e.g. $\hat R \sim \num{E7}$) as the ones of astrophysical interest regarding DM halo application shown in \cref{sec:thermo-stability-analysis}, there are several clockwise turns before the curve start to unwind, thus requiring the same amount of anti-clockwise rotations before the stability is regained.
} From this point and on (which for this rather small $\hat R$ occur at $\theta_0\gtrsim 10$), solutions acquires a \textit{dense core}--\textit{diluted halo} morphology (see solution (3) in \cref{perf_a12}), where the central core is degeneracy-pressure supported (due to Pauli principle) and the atmosphere is thermal pressure supported. For $N < N_{\rm OV}$ this stability trend continue for all accessible values of energy and temperature as $\theta_0$ increases. 

\begin{figure}
     \centering
     \includegraphics[width=\columnwidth]{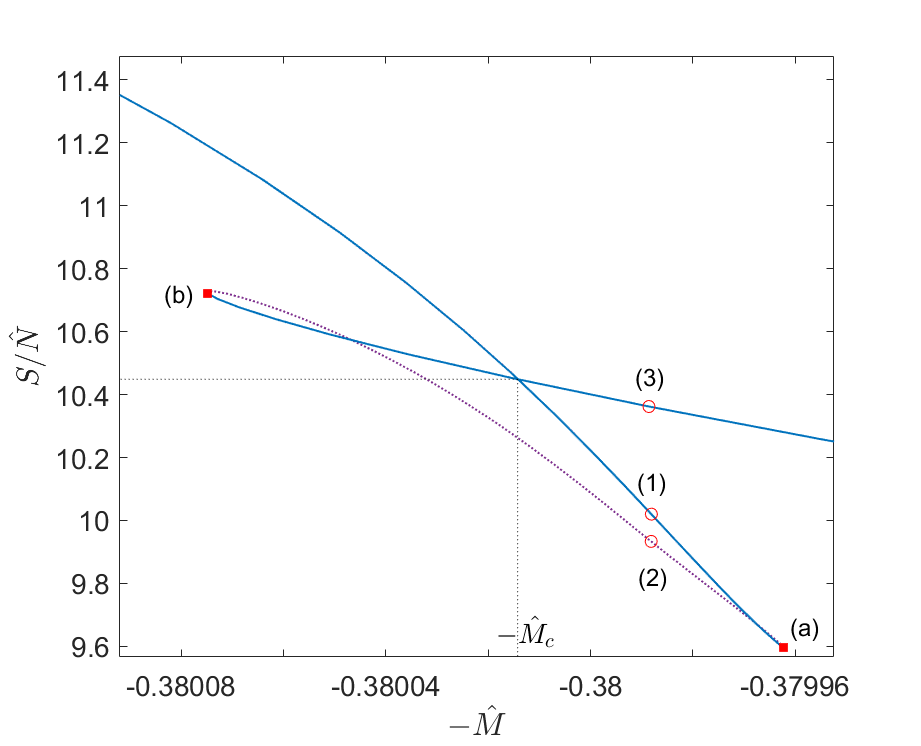}
     \caption{Normalized entropy $S/\hat N$ (according to \cref{eq:Se}) respect to total energy $-\hat M$. Global entropy maxima are clearly distinguished from local entropy maxima, indicating the stable and metastable branches of solutions in \cref{ej_M_c3}. Minimum or saddle points of entropy correspond to the dotted-violet part of the curve, in correspondence with the thermodynamically unstable branch of \cref{ej_M_c3}. At the critical energy $\hat M_c$, a gravitational (first order) phase transition from a gaseous state (1) to a \textit{core-halo} one (3) may take place.}
    \label{S_a10}
\end{figure}

\begin{figure}
     \centering
     \includegraphics[width=\columnwidth]{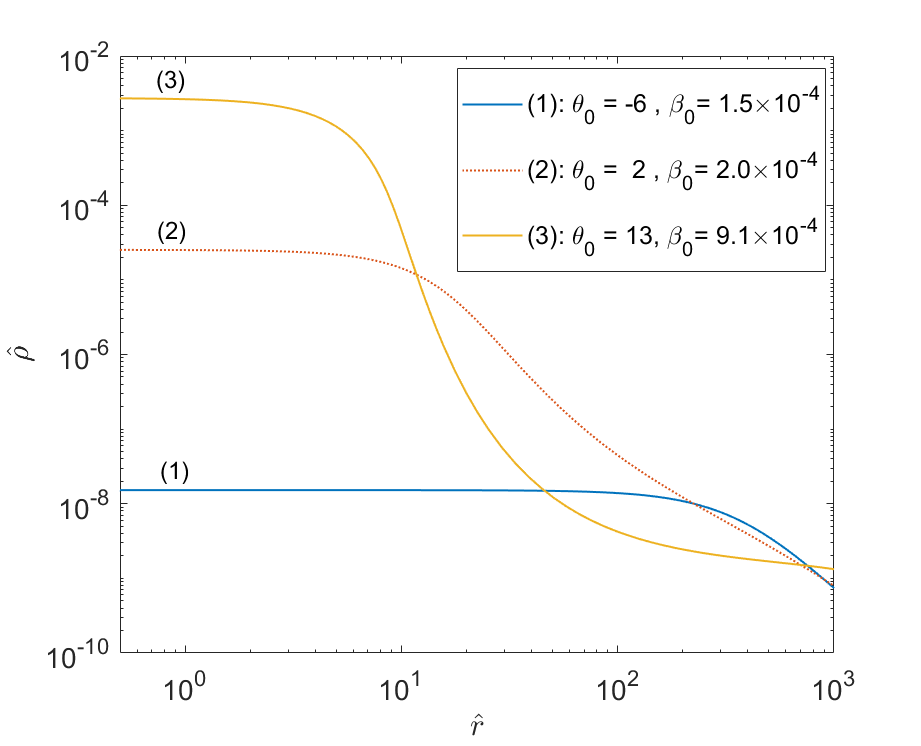}
     \caption{Density profiles corresponding with the equilibrium states of the caloric curve in \cref{ej_M_c3} with energy $\hat M\approx 0.38$. Only the profiles like (1) and the \textit{core-halo} ones like (3) are thermodynamically stable, while profile (2) is thermodynamically unstable. This profiles have no application to DM halos since its boundary value $\hat R$ is orders of magnitude below any realistic halo size.}
    \label{perf_a12}
\end{figure}

Finding stable and metastable solutions via this procedure, equates to proving that such a solution corresponds either to a global or local maximum of the entropy respectively. Unstable solutions correspond either to a minimum or saddle point of entropy. The distinction between different entropy-maxima can be ultimately made by explicitly comparing entropy values for a given energy, as done here in \cref{S_a10} in correspondence with \cref{ej_M_c3}. Metastable fermionic solutions (i.e. local entropy maxima) are of great importance in astrophysics as pointed out in \citet{2005A&A...432..117C}. In particular they are shown to be extremely long-lived and thus reachable in Nature, as proven here as well in \cref{sec:lifetimes} for the case of realistic DM halos in a cosmological setup.


Finally, we make a short description of the gravitational phase-transitions occurring in this kind of self-gravitating systems of fermions at finite $T_\infty$, while we refer the reader to \citet{1999EPJC...11..173B,2006IJMPB..20.3113C,2020EPJB...93..208A} for a broader discussion on the topic (where most of the examples in this Appendix, as the ones displayed in Figs. \ref{ej_M_c3}--\ref{SM_c1}, where already shown). Generally, the phase-transitions (from a gaseous state to another composed by a degenerate core surrounded by a diluted halo) are manifested either in the microcanonical (or canonical) ensembles, through the presence of a multivaluation in the entropy $S$ (or Free Energy $F$) respectively. For relatively small-size systems as studied here ($\hat R=\num{E3}$), we show in \cref{S_a10} the triangle-like shape with the consequent multivaluation in $S$ implying the existence of a (microcanonical) first order phase-transition, occurring at a critical energy of about $\sim 0.5$ in dimensionless units. Interestingly, for smaller size configurations ($\hat R=100$) and similar $\hat N$, we show in \cref{SM_c1} that such a microcanonical phase transition is no longer present, and $S$ increases continuously with energy, while instead, in the canonical ensemble of the same system, the free energy \textit{do show} a multivaluated shape. Indeed, in \cref{F_c1} we see that the free energies of both stable branches equate at the critical temperature $\hat T_c=0.0044$, which indicates that a first order phase transition takes place at that point. Such an apparent discrepancy in the existence (or not) of a gravitational phase-transition is an explicit manifestation of the famous ensemble inequivalence, typical of long-range interaction systems.

\subsection{Lifetime of metastable configurations}
\label{sec:lifetimes}

Due to the long-range nature of the interaction among the self-gravitating particles, it can be shown that the lifetime ($\tau$) of metastable states (i.e. local entropy maxima) scales as ${\rm e}^N$, and therefore they are extremely long-lived for astrophysical systems with large $N$, and cannot be ignored with respect to global entropy maxima states \citep{2005A&A...432..117C}.   

Indeed, in \citet{2005A&A...432..117C} it was explicitly shown that in the microcanonical ensemble, the lifetime of a metastable state can be estimated by $\tau_\mu \sim {\rm e}^{N \Delta s}= {\rm e}^{\Delta S}$. Where $\Delta S= |S_M - S_U|$ is the entropic barrier a system in a metastable state, has to overcome in order to become unstable. \footnote{
    This estimation is valid for any point in the metastable branch, except those with the critical energy corresponding with the microcanonical stability shift \citet{2005A&A...432..117C}, such as point (a).
} While in the canonical ensemble, the corresponding lifetime scales as $\tau_c \sim {\rm e}^{N \Delta j}= {\rm e}^{\Delta F}$, where $\Delta F= |F_M - F_U|$ is the free energy barrier between the same two states as before and $j$ the free energy per particle. Interestingly, in \citet{2005A&A...432..117C} it was possible to obtain (from a stochastic approach based on a dynamical model for self-gravitating Brownian particles) a first-principle justification of the full life-time formula, which in the case of the canonical ensemble reads,

\begin{equation}
    \label{tlf}
    \tau_c=\pi(k_BT)^2\frac{\sqrt{C_U\lvert C_M\rvert}}{D}{\rm e}^{(F_M-F_U)/k_BT},
\end{equation}
with $C_M$, $C_U$ the specific heats of the metastable and unstable states, $k_B$ is the Boltzmann constant, $T$ the fixed temperature corresponding with the free energy barrier, and $D$ is the diffusion coefficient of the metastable configuration which in a mean field approximation is $D=3M^2c^2/N$. Therefore, the full caloric curve together with the two (metastable and unstable) selected states at a given temperature (or energy for the microcanonical ensemble) are enough, in order to calculate $\tau$.

\begin{figure}
     \centering
     \includegraphics[width=0.95\columnwidth]{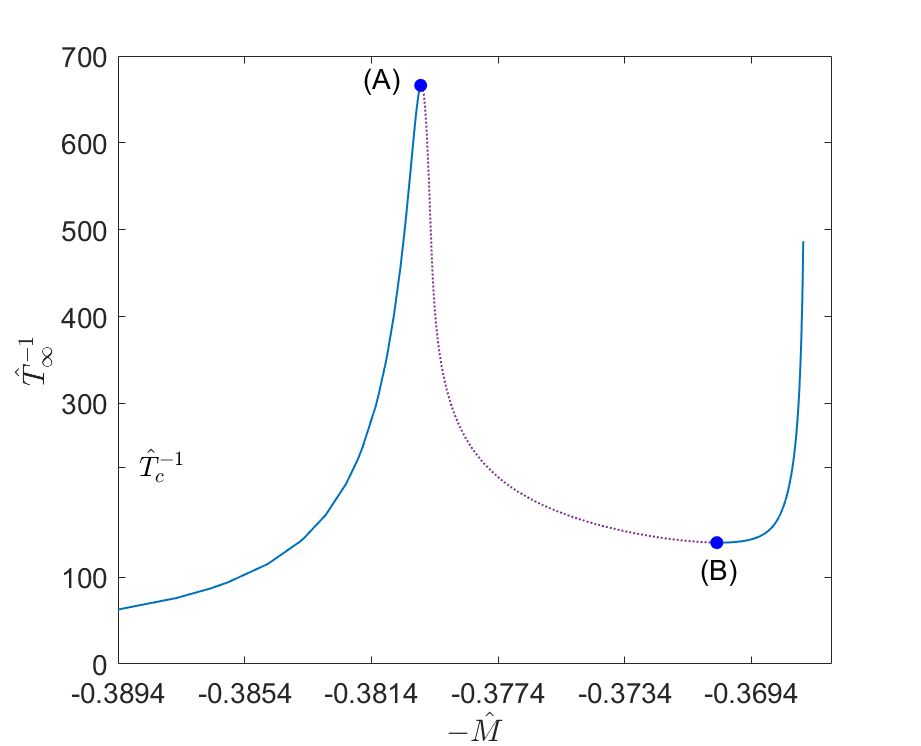}
     \caption{Series of equilibrium solutions along the caloric curve for box-confined configurations of fermions with $\hat N=0.38 < N_{\rm OV}$ and $\hat R=10^2$. According to the Katz criterion (canonical ensemble), at the maximum (A) of the caloric curve, it losses a stability mode, which is re-gained at the minimum (B) of the curve. The stable (unstable) branches in relation with the minimum (maximum/saddle points) of free energy is shown in \cref{F_c1}. For such a low boundary size $\hat R$ the curve develops an N-shape showing no microcanonical phase-transition at all (at difference with \cref{ej_M_c3}).}
    \label{M_c1}
\end{figure}

\begin{figure}
     \centering
     \includegraphics[width=0.95\columnwidth]{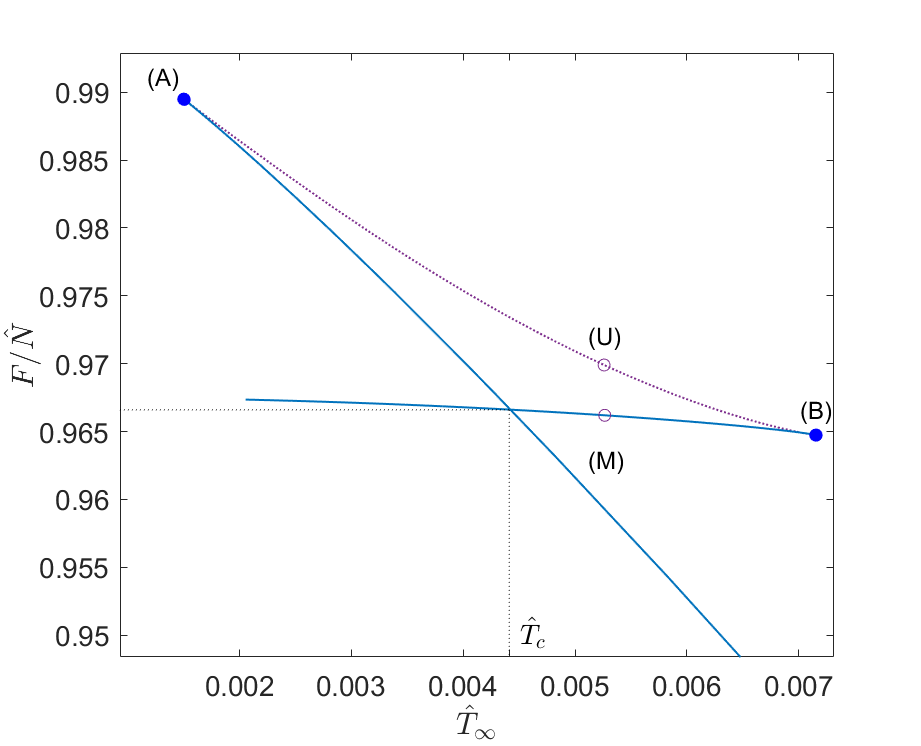}
     \caption{Normalized free energy $F/\hat N$ (according to \cref{eq:Fp}) respect to dimensionless temperature $\hat T_\infty$. The critical temperature $\hat T_\infty$ indicates the place in the caloric curve \cref{M_c1} where a (first order) canonical phase transition may occur. This result was first found in \citet{1999EPJC...11..173B}.}
    \label{F_c1}
\end{figure}

\begin{figure}
     \centering
     \includegraphics[width=0.95\columnwidth]{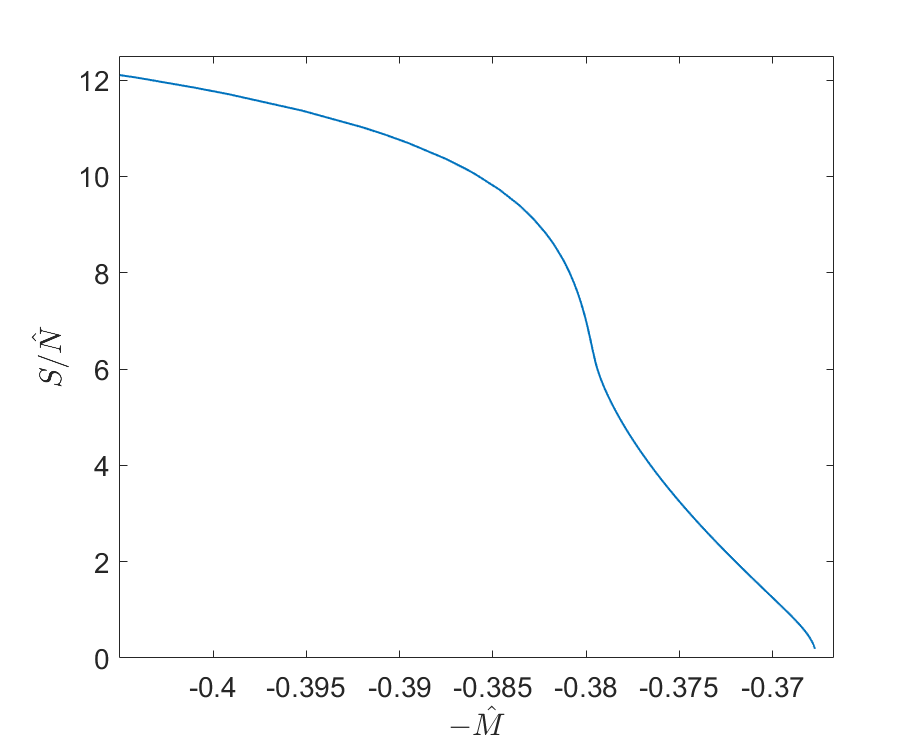}
     \caption{Normalized entropy $S/\hat N$ respect to dimensionless energy $-\hat M$ corresponding to the caloric curve of \cref{M_c1}. The continuous (not multi-valuated) trend of the entropy, indicates that no microcanonical phase transition is present at difference with the canonical ensemble (\cref{F_c1}), evidencing the ensemble-inequivalence typical of long-range interaction systems.}
    \label{SM_c1}
\end{figure}

Next, we provide the explicit calculations of the life-times of the metastable states in different scenarios: first we give the value of $\tau_c$ from \cref{tlf} in terms of the caloric curve given in \cref{M_c1}, as a pedagogical example. We do it under the choice of $\hat R=100$, $\hat N=0.38$ (i.e. not applicable to DM halos), for particular unstable and metastable states (labelled with (U) and (M) respectively) with temperature  $\hat T=0.0053$ (or $T=\SI{6.4E5}{K}$) as shown in \cref{F_c1}. This gives as a result a tremendously long-lived metastable state with $\tau_c = \num{7.8E-28}\,{\rm e}^{\num{4.2E71}}\,\si{\second}$, which can be considered as $\infty$, being $\Delta j=4.2\times 10^{-3}$ and $N=10^{74}$ fermions. 

Finally, we estimate the lifetimes $\tau_\mu \sim {\rm e}^{N \Delta s}$ of the metastable states selected in the three cases analysed within the microcanonical ensemble in \cref{sec:thermo-stability-analysis}. They are labelled as (5) for the case of \cref{fig:calcurvebox1}, and (3) in the cases of \cref{Ma16} and \cref{Mk24}, with the corresponding unstable ones forming the entropic barrier as mentioned above. We first calculate the barrier of entropy per particle $\Delta s$ between the selected states using formula \cref{eq:Se}, and then multiply by $N$. Interestingly, in all the three cases of astrophysical interest studied here $\Delta s\gtrsim 0.2$ (rising up to $\sim 10^3$ in the case of \cref{Ma16}, while reaching a much lower value of $0.24$ in \cref{Mk24} given the rather close position between the given unstable and metastable states), and $N\sim \mathcal{O}(10^{70})$, thus making the lifetimes $\tau_\mu$ of these metastable states essentially infinite for all the cases here studied.

%Important features: no phase transition nor metastable states, just stable/unstable solutions, no explosion nor collapse; minimal temperature; spiral structure (BH). Microcanical: identical to previous case. \todo[inline]{This paragraph looks unfinished, just a collection of notes.}

\section{DM halo formation within a WDM paradigm}
\label{sec:appendix:halo-formation}

%%%%%%%%%%%%%
%Formula para M(R)
%%%%%%%%%%%%%

%Por que virialization
A thermodynamical stability analysis of a quasi-relaxed system of self-gravitating fermions has been performed in the microcanonical ensemble within GR in \cref{sec:thermo-stability-analysis}. As explained in the introduction, violent-relaxation is the main underlying relaxation process able to lead the fermionic halo into the steady we observe, a process occurring at late stages on non-linear structure formation. 
%(NOT SO SURE - Check further Press-Schechter and subsequent violent relaxation from Chavanis but also Tremaine and Lynden-Bell)
%We are interested in obtaining conclusions for these systems right after the virialization process, and before (as much as possible) the system undergoes mergers: these processes may modify some of the assumptions of the thermodynamic analysis, namely constant energy and particle number.
The boundary conditions for the thermodynamical analysis of these self-gravitating systems, e.g. particle number and total radius, are valid after the virialization of the structure in the context of hierarchical structure formation. Given a cosmological evolution model, the Press-Schechter (PS) theory \citep{PressSchechter74} or its subsequent extensions (see e.g. \citealp{MoWhiteBook}) can be used to get the relevant astrophysical magnitudes at formation such as virial mass and radius at a given collapse redshift $z^{*}$ \textbf{($z^{*}\equiv z_{vir}$)}.
%trace the formation and evolution of these DM halos. 
%M(R,t0) es facil pero M(R,t) no tanto
Relating the mass of a given DM halo with its spatial extent follows directly from the definitions of a virial radius and the background density of the universe. However, estimating a typical redshift in which the structure is formed becomes key to the analysis, as the density is a time dependent quantity.
%Queremos sacar M(R,t)
So, our objective for this appendix is, given a (total) halo mass scale $M \equiv M_{vir}$, to obtain an approximation to the spatial extent of the system right after virialization is complete: $r_{\rm vir}$ measured at the most probable gravitational collapse time $t_{\rm vir}$. 
%Paso a paso
We use the Press-Schechter formalism to obtain a good estimation on the most probable collapse time for a given mass scale $t_{\rm PS}(M)$. Then, we use this estimate to obtain the virial radius of the system right after it virialized $r_{\rm vir}(M, t_{\rm PS}(M))$.

A typical  definition for the boundary of a halo is set by the `virial radius' $r_{200}$: it is defined so that the mean density of the halo within this radius is $200$ times the critical density. The mass inside $r_{200}$ is used as a measure of the total mass of the halo, and is related to $r_{200}$ \citep{2008gady.book.....B}
% ,VanDenBosch13

\begin{equation}
M_{vir} = 200 \frac{4}{3} \pi r_{200}^3 \rho_c (t) = 100 \frac{H_0^2 r_{200}^3 (1+z(t))^3 }{G}  \ ,
\label{eq:SecPS_M200}
\end{equation}

%%%%%%%%%%%%%
%Justificacion: Colapso esferico y radio virial
%%%%%%%%%%%%%

\noindent where $\rho_c$ is the critical density of the universe, $H_0$ is the Hubble constant, $z$ is the redshift, $G$ the gravitational constant and we have taken a flat, matter dominated universe for simplicity. This overdensity value is motivated by the spherical collapse model, which suggests that regions in which the background density exceeds approximately this value should be part of a virialized halo \citep{MoWhiteBook,2008gady.book.....B}.

%%%%%%%%%%%%%
%M(z) en WDM?
%%%%%%%%%%%%%

In order to define a characteristic collapse redshift $z^{*}$ for a given mass scale $M$ we turn to the Press-Schechter formalism \citep{PressSchechter74}, that provides a way of understanding how nonlinear collapsed structures form in a hierarchical way. According to this model, halos with mass $M$ can only form in a significant number when the mass variance $\sigma(M)$, defined as \citep{MoWhiteBook,2008gady.book.....B}

\begin{equation}
\sigma^2(M) = \frac{1}{2\pi^2} \int_0^{\infty} P(k) W^2(k,R) k^2 {\rm d}k \ ,
\end{equation}  

\noindent exceeds a critical value $\delta_c(t)=1.69/D(t)$ given by spherical collapse, where $D(t)$ is the linear growth rate of perturbations, $P(k)$ is the matter power spectrum and $W(k,R)$ is a window function of characteristic radius $R$ (taken as a \textit{top hat function} here, see e.g. \citealp{MoWhiteBook}). This radius $R$ is the lagrangian radius corresponding to a mass scale $M$. Thus, we can define a characteristic collapse mass $M^{*}(z)$ by

\begin{equation}
\sigma (M^{*}) = \delta_c (t) \ .
\end{equation}

\begin{figure}
     \centering
     \includegraphics[width=\columnwidth]{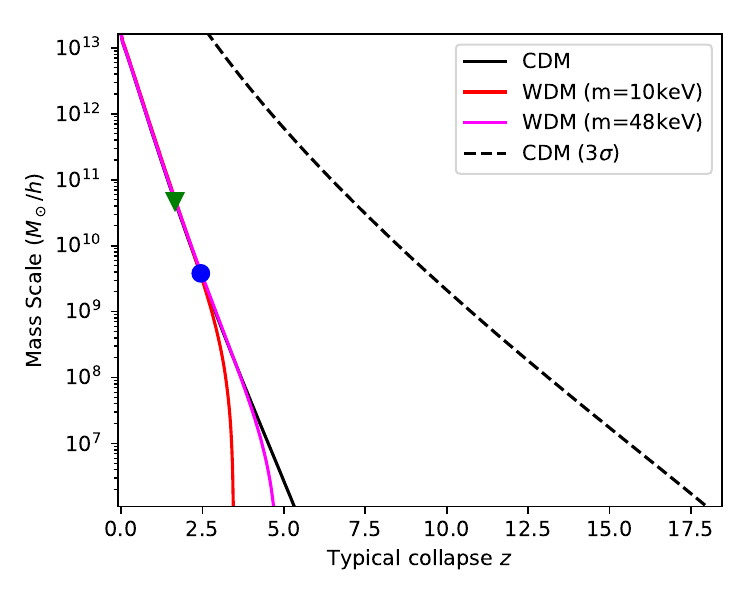}
     \caption{Characteristic collapse redshift $z^*$ as a function of substructure mass $M_{vir}$. CDM models in black line and WDM thermal relic models with $mc^2 = \SI{10}{\kilo\eV}$ and $mc^2 = \SI{48}{\kilo\eV}$ in red and magenta, respectively. An additional dotted line represents the collapse redshift of 3$\sigma$ overdensities in CDM. The mass - virial radius combinations used for DM halo models in \cref{sec:thermo-stability-analysis} are marked on the plot with a green triangle and a blue dot.} 
    \label{fig:SecPS_Z_M}
\end{figure}

So at redshift $z$ halos are formed in significant numbers for masses $M \leq M^{*}$. We can invert this relation so that, for a given substructure mass $M$, we can obtain a typical collapse redshift $z^* (M)$: this relation can be seen in \cref{fig:SecPS_Z_M}, together with two selected values of $M_{vir}=6.2\times10^9 M_\odot$ (blue dot), and $M_{vir}=5.9\times10^{10} M_\odot$ (green triangle) for a WDM cosmology with $mc^2 = \SI{10}{\kilo\eV}$, corresponding with the values used in \cref{sec:box-case}. The value of $M_{vir}=5.4\times10^{10} M_\odot$, at $z^*=1.7$ in the case of a WDM cosmology with $mc^2 = \SI{48}{\kilo\eV}$ (not displayed in the plot), is used in \cref{sec:cutoff-case} and in \cref{Sigma-T}. 
%[RAFA DISCUSS a bit thedifference between dashed and continuous curve in this fig., mentioning the expected redshift of about 10 at early DM halo formation for typical average-mass halos.]
As the collapse mass $M^{*}$ indicates the $z$ in which \textit{most} structures of such mass are collapsed, we also plot in \cref{fig:SecPS_Z_M} the $3\sigma$ collapse mass defined as $3 \sigma (M^{*}_{3\sigma}) = \delta_c (t)$ to indicate the expected formation redshift of the earliest halos (see e.g. \cite{2008gady.book.....B}). For average to low-mass halos such as the ones considered in \cref{sec:box-case} and \cref{sec:cutoff-case}, the PS formalism expects early halo formation at a redshift up to $z\approx 10$ as shown in \cref{fig:SecPS_Z_M}.

\begin{figure}
     \centering
     \includegraphics[width=\columnwidth]{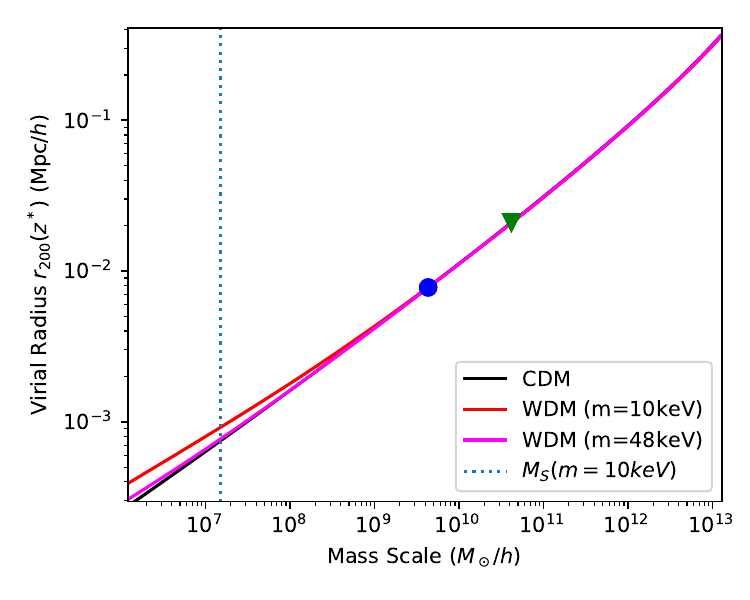}
     \caption{Virial radius $r_{200}$ at virialization time as a function of substructure mass $M_{vir}$. CDM models in black line and WDM thermal relic models with $mc^2 = \SI{10}{\kilo\eV}$ and $mc^2 = \SI{48}{\kilo\eV}$ in red and magenta, respectively. In blue dotted line, the free streaming mass scale $M_{S} = \SI{2E7}{\Msun}$. The green triangle and blue dot correspond to the ones selected in \cref{fig:SecPS_Z_M} for the astrophysical application of \cref{sec:thermo-stability-analysis}.} 
    \label{fig:SecPS_MR200_MFS}
\end{figure}

Once this typical collapse redshift is obtained, it is possible to evaluate the relation between scale mass and radius \eqref{eq:SecPS_M200}, using $z^* (M)$ to obtain \textit{an estimation of the virial radius of substructure of mass $M$, evaluated at the time in which most of this substructure is already collapsed}. This is shown in \cref{fig:SecPS_MR200_MFS} for the above selected values of $M_{vir}$, implying $r_{\rm vir}\equiv r_{200}=11.1$ kpc (blue dot), $r_{vir}=29.7$ kpc (green triangle, for a $mc^2 = \SI{10}{\kilo\eV}$ cosmology), and $r_{vir}=28.5$ kpc (for a $mc^2 = \SI{48}{\kilo\eV}$ cosmology, not displayed in the plot).

%%%%%%%%%%%%
%Parametros usados en modelos WDM
%%%%%%%%%%%%

In \cref{fig:SecPS_MR200_MFS} and \cref{fig:SecPS_Z_M} the results obtained here using several different cosmological models, are compared. For $\Lambda$CDM models we simulate the power spectrum $P(k)$ using best fit parameters from Planck's 2018 data release \citep{2020A&A...641A...6P}, and compare these results with the ones obtained by replacing CDM for sterile neutrino WDM components of $\SI{10}{\kilo\eV}$ and $\SI{48}{\kilo\eV}$, leaving all other cosmology parameters equal. In the later, the power spectra is simulated using CLASS version 2.7.2 \citep{2011JCAP...09..032L} and sterile neutrino WDM phase space distribution at production simulated using \cite{Venumadhav} with mixing angles of $\theta^2 = \num{5E-11}$ and $\theta^2 = \num{E-12}$ for the $\SI{10}{\kilo\eV}$ and $\SI{48}{\kilo\eV}$ models respectively. We can see in \cref{fig:SecPS_MR200_MFS} that differences between CDM and WDM models are insignificant with respect to this relation.

%%%%%%%%%%%%%
%WDM: Escala de cutoff basada en Venumadhav 2015
%%%%%%%%%%%%%

Also plotted in \cref{fig:SecPS_MR200_MFS} we can see the corresponding free streaming scale of the $mc^2=\SI{10}{\kilo\eV}$ WDM model, indicating the smallest non-suppressed structures in the power spectrum. Typically, WDM models establish a cutoff in the power spectrum of metric perturbations due to free-steaming of particles. This implies that the formation of objects under a certain length scale is suppressed and the number of small, low mass halos and satellite galaxies is significantly lower. We plot an estimate to this mass scales using the prescription in \cite{VielLesgourgues} for sterile neutrino free streaming.

\bsp	% typesetting comment
\label{lastpage}
\end{document}